%% file: main.tex
\documentclass[ALICE,manyauthors]{cernphprep}
\usepackage[comma,square,numbers,sort&compress]{natbib}
\usepackage{hyperref}
\usepackage{lineno}
\usepackage[T1]{fontenc}
\usepackage{xspace}
\usepackage[utf8]{inputenc}
\usepackage{booktabs}
\usepackage{multirow}
\usepackage{color}
\usepackage{euscript}
\usepackage{rotating}
\usepackage[dvipsnames]{xcolor}

\usepackage{silence}
\usepackage[T1]{fontenc}
\usepackage{orcidlink}

\begin{document}
\input{commands.tex}

\begin{titlepage}
\PHyear{2024}       
\PHnumber{243}      
\PHdate{18 September}  

\title{First observation of strange baryon enhancement with effective energy in pp collisions at the LHC}
\ShortTitle{Strange baryon enhancement with effective energy in pp collisions}   

\Collaboration{ALICE Collaboration\thanks{See Appendix~\ref{app:collab} for the list of collaboration members}}
\ShortAuthor{ALICE Collaboration} 

\begin{abstract}
The production of (multi-)strange hadrons is measured at midrapidity in proton--proton collisions at $\s = 13 \rm ~TeV$ as a function of the local charged-particle multiplicity in the pseudorapidity interval ${|\eta|<0.5}$ and of the very-forward energy measured by the ALICE Zero-Degree Calorimeters.
The latter provides information on the effective energy, i.e. the energy available for particle production in the collision once subtracted from the centre-of-mass energy.
The yields of \kzero, $\lmb+\almb$, and $\Xi^{-}+\overline{\Xi}^{+}$ per charged-particle increase with the effective energy. In addition, this work exploits a multi-differential approach to decouple the roles of local multiplicity and effective energy in such an enhancement.
The results presented in this article provide new insights into the interplay between global properties of the collision, such as the initial available energy in the event, and the locally produced final hadronic state, connected to the charged-particle multiplicity at midrapidity.
Notably, a strong increase of strange baryon production with effective energy is observed for fixed charged-particle multiplicity at midrapidity. These results are discussed within the context of existing phenomenological models of hadronisation implemented in different tunes of the PYTHIA 8 event generator.

\end{abstract}
\end{titlepage}

\setcounter{page}{2} 

\input{Chapters/introduction}
\input{Chapters/experiment}
\input{Chapters/analysis}

\input{Chapters/systematics}

\input{Chapters/results}

\input{Chapters/conclusions}



\newenvironment{acknowledgement}{\relax}{\relax}
\begin{acknowledgement}
\section*{Acknowledgements}
\input{fa_2024-08-27_Opt_C.tex}
\end{acknowledgement}

\bibliographystyle{utphys}   

\bibliography{bibliography}

\newpage
\appendix

%
%

\section{The ALICE Collaboration}
\label{app:collab}
\input{Alice_Authorlist_2024-08-27_Opt_C.tex}
\end{document}

%% file: commands.tex
%

\newcommand{\pp}           {pp\xspace}
\newcommand{\ppbar}        {\mbox{$\mathrm {p\overline{p}}$}\xspace}
\newcommand{\XeXe}         {\mbox{Xe--Xe}\xspace}
\newcommand{\PbPb}         {\mbox{Pb--Pb}\xspace}
\newcommand{\pA}           {\mbox{pA}\xspace}
\newcommand{\pPb}          {\mbox{p--Pb}\xspace}
\newcommand{\AuAu}         {\mbox{Au--Au}\xspace}
\newcommand{\dAu}          {\mbox{d--Au}\xspace}

\newcommand{\s}            {\ensuremath{\sqrt{s}}\xspace}
\newcommand{\snn}          {\ensuremath{\sqrt{s_{\mathrm{NN}}}}\xspace}
\newcommand{\pt}           {\ensuremath{p_{\rm T}}\xspace}
\newcommand{\meanpt}       {$\langle p_{\mathrm{T}}\rangle$\xspace}
\newcommand{\ycms}         {\ensuremath{y_{\rm CMS}}\xspace}
\newcommand{\ylab}         {\ensuremath{y_{\rm lab}}\xspace}
\newcommand{\etarange}[1]  {\mbox{$\left | \eta \right |~<~#1$}}
\newcommand{\yrange}[1]    {\mbox{$\left | y \right |~<~#1$}}
\newcommand{\dndy}         {\ensuremath{\mathrm{d}N_\mathrm{ch}/\mathrm{d}y}\xspace}
\newcommand{\dndeta}       {\ensuremath{\mathrm{d}N_\mathrm{ch}/\mathrm{d}\eta}\xspace}
\newcommand{\avdndeta}     {\ensuremath{\langle\dndeta\rangle}\xspace}
\newcommand{\dNdy}         {\ensuremath{\mathrm{d}N_\mathrm{ch}/\mathrm{d}y}\xspace}
\newcommand{\Npart}        {\ensuremath{N_\mathrm{part}}\xspace}
\newcommand{\Ncoll}        {\ensuremath{N_\mathrm{coll}}\xspace}
\newcommand{\dEdx}         {\ensuremath{\textrm{d}E/\textrm{d}x}\xspace}
\newcommand{\RpPb}         {\ensuremath{R_{\rm pPb}}\xspace}

\newcommand{\nineH}        {$\sqrt{s}~=~0.9$~Te\kern-.1emV\xspace}
\newcommand{\seven}        {$\sqrt{s}~=~7$~Te\kern-.1emV\xspace}
\newcommand{\twoH}         {$\sqrt{s}~=~0.2$~Te\kern-.1emV\xspace}
\newcommand{\twosevensix}  {$\sqrt{s}~=~2.76$~Te\kern-.1emV\xspace}
\newcommand{\five}         {$\sqrt{s}~=~5.02$~Te\kern-.1emV\xspace}
\newcommand{\twosevensixnn}{$\sqrt{s_{\mathrm{NN}}}~=~2.76$~Te\kern-.1emV\xspace}
\newcommand{\fivenn}       {$\sqrt{s_{\mathrm{NN}}}~=~5.02$~Te\kern-.1emV\xspace}
\newcommand{\LT}           {L{\'e}vy-Tsallis\xspace}
\newcommand{\GeVc}         {Ge\kern-.1emV/$c$\xspace}
\newcommand{\MeVc}         {Me\kern-.1emV/$c$\xspace}
\newcommand{\TeV}          {Te\kern-.1emV\xspace}
\newcommand{\GeV}          {Ge\kern-.1emV\xspace}
\newcommand{\MeV}          {Me\kern-.1emV\xspace}
\newcommand{\GeVmass}      {Ge\kern-.2emV/$c^2$\xspace}
\newcommand{\MeVmass}      {Me\kern-.2emV/$c^2$\xspace}
\newcommand{\lumi}         {\ensuremath{\mathcal{L}}\xspace}

\newcommand{\ITS}          {\rm{ITS}\xspace}
\newcommand{\TOF}          {\rm{TOF}\xspace}
\newcommand{\ZDC}          {\rm{ZDC}\xspace}
\newcommand{\ZDCs}         {\rm{ZDCs}\xspace}
\newcommand{\ZNA}          {\rm{ZNA}\xspace}
\newcommand{\ZNC}          {\rm{ZNC}\xspace}
\newcommand{\SPD}          {\rm{SPD}\xspace}
\newcommand{\SDD}          {\rm{SDD}\xspace}
\newcommand{\SSD}          {\rm{SSD}\xspace}
\newcommand{\TPC}          {\rm{TPC}\xspace}
\newcommand{\TRD}          {\rm{TRD}\xspace}
\newcommand{\VZERO}        {\rm{V0}\xspace}
\newcommand{\VZEROA}       {\rm{V0A}\xspace}
\newcommand{\VZEROC}       {\rm{V0C}\xspace}
\newcommand{\Vdecay} 	   {\ensuremath{V^{0}}\xspace}

\newcommand{\ee}           {\ensuremath{e^{+}e^{-}}} 
\newcommand{\pip}          {\ensuremath{\pi^{+}}\xspace}
\newcommand{\pim}          {\ensuremath{\pi^{-}}\xspace}
\newcommand{\kap}          {\ensuremath{\rm{K}^{+}}\xspace}
\newcommand{\kam}          {\ensuremath{\rm{K}^{-}}\xspace}
\newcommand{\pbar}         {\ensuremath{\rm\overline{p}}\xspace}
\newcommand{\kzero}        {\ensuremath{{\rm K}^{0}_{\rm{S}}}\xspace}
\newcommand{\lmb}          {\ensuremath{\Lambda}\xspace}
\newcommand{\almb}         {\ensuremath{\overline{\Lambda}}\xspace}
\newcommand{\lmbs}         {\ensuremath{\Lambda+\overline{\Lambda}}\xspace}
\newcommand{\Om}           {\ensuremath{\Omega^-}\xspace}
\newcommand{\Mo}           {\ensuremath{\overline{\Omega}^+}\xspace}
\newcommand{\X}            {\ensuremath{\Xi^-}\xspace}
\newcommand{\Ix}           {\ensuremath{\overline{\Xi}^+}\xspace}
\newcommand{\Xis}          {\ensuremath{\X+\Ix}\xspace}
\newcommand{\Xiz}          {\ensuremath{\Xi^{0}}\xspace}
\newcommand{\Oms}          {\ensuremath{\Omega^{\pm}}\xspace}
\newcommand{\degree}       {\ensuremath{^{\rm o}}\xspace}

%% file: Chapters/introduction.tex
\section{Introduction}
\label{sec:introduction}

One of the main challenges of hadron physics is the understanding of the origin of strangeness enhancement in high-energy hadronic collisions. This prominent phenomenon consists in the continuous increase of the strange to non-strange hadron yield ratios with increasing charged-particle pseudorapidity density at midrapidity (\dndeta), from low-multiplicity pp collisions, characterised by \dndeta $\sim$ 3 (approximately 40$\%$ of the minimum-bias value), up to high-multiplicity p--Pb collisions, with \dndeta $\sim$ 50~\cite{strangenessEnhancementNature,strangenessEnhancementpp,strangeness_pPb1,strangeness_pPb2,strangeness_PbPb}. For higher values of charged-particle multiplicity, the yield ratios stay approximately constant up to the highest multiplicities reached in central Pb--Pb collisions (\dndeta $\sim$ 2000)~\cite{strangeness_PbPb}.
The enhanced production of strange hadrons in heavy-ion collisions compared to minimum-bias pp collisions, first observed at the SPS experiments in the late 90's~\cite{1999401}, was historically considered one of the signatures of quark--gluon plasma (QGP) formation~\cite{PhysRevLett.48.1066}. The observation of a smooth evolution of the strange to non-strange yield ratios across different collision systems and centre-of-mass energies suggests a continuous transition of the underlying hadronisation mechanism that determines the hadron chemistry, i.e.~the relative abundances of different hadron species, in high-energy hadronic collisions.

Important insights into the origin of strangeness production in small collision systems came from the results of the ALICE Collaboration on strange hadron production associated with hard scattering processes and to the underlying event (UE) in pp and p--Pb collisions~\cite{strangeness_inJet1,strangeness_inJet2}.
In the cited ALICE results, in particular, different values and \pt dependencies of strange baryon-to-meson and baryon-to-baryon ratios are measured in- and out-of-jets. The baryon-to-meson ratio shows an increasing trend at low \pt, reaching a maximum value at $\pt \sim 3$ GeV/$c$ and then decreasing towards higher \pt, giving rise to a broad peak at intermediate \pt. This peak is more pronounced for strange baryons and mesons produced in the UE, while a lower ratio and a milder evolution with \pt are observed within jets.
The study of strange hadron production in hard and soft quantum chromodynamics (QCD) regimes was recently extended using the two-particle correlation technique as a function of the charged-particle multiplicity~\cite{PaperChiaraDM,ALICE:2024aid,ALICE:2024iqc}.
Strange hadrons are found to be mainly produced in the direction transverse to the leading particle, which is dominated by the UE, where the yields increase significantly with \dndeta.
Instead, a weaker multiplicity dependence is observed for the production within a rectangular region around the leading particle, a proxy for the jet axis.
Strangeness enhancement with multiplicity, studied through the ratio of multi-strange baryons to strange mesons, was observed in both the toward and transverse-to-leading regions with proportional slopes, suggesting it to be a common feature of both particle production regimes.
This effect was further investigated by studying the event as a whole, selecting isotropic collisions, supposedly driven by large underlying events, and events characterised by jetty typologies. An enhancement of strange hadrons is observed in collisions characterised by an isotropic topology with respect to events with a jet-like topology~\cite{spherocity}.
Despite a substantial body of high-quality experimental results, a comprehensive understanding of the underlying mechanisms of strangeness production in small systems remains unclear, stressing the need for further investigation.

This work exploits a new approach to study strangeness production in pp collisions.
For the first time, the concept of effective energy ($E_{\rm eff}$) is introduced in hadronic interactions at the Large Hadron Collider (LHC).
In pp collisions, the effective energy is defined as the energy available for particle production, which is reduced with respect to the nominal centre-of-mass energy due to the emission of leading baryons at very-forward rapidity (leading-baryon effect)~\cite{Zichichi2}.
The effective energy was extensively studied by past experiments at the CERN ISR by investigating its correlation with collision event properties such as the charged-particle multiplicity, to shed light on the universal features of the QCD.
In particular, the results from several past experiments show that the value of the charged-particle multiplicity at a given centre-of-mass energy in the case of pp ($\rm p\overline{p}$) collisions is systematically lower than in $\rm e^{-}e^{+}$ data at the same energy.
However, the outcome of the studies performed at the ISR showed that a universal dependence can be observed if the appropriate definition of the energy available for particle production (effective energy) is used~\cite{Basile1,
BCFlast,
momBCF,
Basile_1,
Basile_2,
Basile3,
finalpp,
Basile4}.
One way of estimating the effective energy is by measuring the energy of the leading baryons ($E_{\rm leading}$) produced at forward rapidities in each event hemisphere.
The ALICE experiment is well suited to measure the energy deposited in the very-forward region, which is expected to be mainly due to baryons with kinematics close to the beam rapidity~\cite{Akindinov_2007}.
This observable can be directly connected to the leading-baryon effect, related to the baryon number conservation of incident hadrons.
Assuming the energy deposited at forward rapidity provides an estimation of the energy of leading baryons, this provides an indirect measurement of the effective energy:
\begin{equation}
E_{\rm eff} = \sqrt{s} - E_{\rm leading} \approx \sqrt{s} - E_{\rm forward}  \quad .
\end{equation}

Several phenomenological approaches traditionally employed to model hadronic interactions have been investigated to probe the origin of strangeness enhancement in small collision systems, including a statistical hadronisation description using the canonical suppression approach~\cite{CSM_PRC_2019}, rope hadronisation models including colour reconnection (CR) effects~\cite{CR_PRD_2019}, and two-component (core--corona) models~\cite{CC_PRC_2020}.
The general-purpose Monte Carlo generator PYTHIA~\cite{pythia8generic}, in particular, implements colour string fragmentation at its core, and features an intrinsic correlation between (multi-)strange hadron production and the number of multiparton interactions (MPIs) in pp collisions, which are directly connected to the final-state charged-particle multiplicity at midrapidity.
As shown in Ref.~\cite{correlationZDCdNdeta}, there is an anti-correlation between the multiplicity of charged-particles measured at midrapidity and the energy measured at very-forward rapidity.
Within the context of these models, the forward energy is indicative of the number of MPIs.
In particular, a decrease in the average forward energy is predicted for an increasing number of MPIs.
Indeed, the effective energy available for particle production at midrapidity is expected to be strongly correlated with the number of parton--parton collisions that occurred in the event.

In this paper, strangeness enhancement is studied by measuring \kzero, \lmbs, and $\Xi^{-}+\overline{\Xi}^{+}$ in double-differential classes as a function of the charged-particle pseudorapidity density at midrapidity (\dndeta) and the very-forward energy (as a proxy for the leading energy).
The ratio of strange hadron yields per charged-particle is related to the forward energy deposit for similar average \dndeta values and vice versa. This novel experimental technique is used to test the traditional paradigm in which strangeness enhancement is found to increase with midrapidity multiplicity.
Such an approach allows for ideal decoupling of the interplay, for strange hadron production, between global properties of the collision and the produced final hadronic state, under the assumption that midrapidity multiplicity and leading energy are independent proxies, given the large $\eta$ separation.

The paper is organised as follows.~Section~\ref{sec:Experiment} presents the ALICE experimental apparatus, and Sec.~\ref{sec:analysis} and~\ref{subsec:AnalysisStrategy} discuss the data set and the analysis techniques, respectively. Section~\ref{subsec:Phenomodels} presents the studies performed on the phenomenological models, Sec.~\ref{subsec:CascadeAndV0Selection} outlines the techniques used to reconstruct strange hadrons, and Sec.~\ref{sec:SystematicUncertainties} covers the evaluation of systematic uncertainties. Finally, Sec.~\ref{sec:results} presents the results and Sec.~\ref{sec:conclusions} reports the conclusions.

%% file: Chapters/experiment.tex
\section{Experimental apparatus}
\label{sec:Experiment}

ALICE is a general-purpose experiment at the LHC dedicated to the study of ultra-relativistic hadronic collisions. A detailed description of the ALICE apparatus and its performance can be found in Refs.~\cite{ALICEexperiment} and~\cite{ALICEperformance}. In the following, only the sub-detector systems used for the analysis presented in this paper are described.

Trajectories of charged particles are reconstructed in the ALICE central barrel with the Inner Tracking System (ITS)~\cite{ITS} and the Time Projection Chamber (TPC)~\cite{TPC}. These sub-detectors are located within a large solenoidal magnet, providing a highly homogeneous magnetic field of 0.5 T parallel to the beam axis.
The ITS used during the LHC Run 2 consisted of six cylindrical layers of silicon detectors, concentric and coaxial to the beam pipe, with a total pseudorapidity coverage $|\eta| < 0.9$ with respect to the nominal interaction point. Three different technologies were used for this detector: the two innermost layers consisted of silicon pixel detectors (SPD), the two central layers of silicon drift detectors (SDD), and the two outermost layers of double-sided silicon strip detectors (SSD). The ITS is used in the determination of primary and secondary vertices, and in the track reconstruction.
For the purpose of this analysis the first two ITS layers were used to provide a midrapidity multiplicity estimator independent of the ITS track reconstruction.

The TPC is the largest detector in the ALICE central barrel, with a pseudorapidity coverage $|\eta| < 0.9$. It is used for charged-particle track reconstruction, momentum measurement, and particle identification (PID) via the measurement of the specific energy loss ($\mathrm{d}E/\mathrm{d}x$) of particles in the TPC gas. The TPC provides up to 159 spatial points per track for charged-particle reconstruction. The \dEdx resolution depends on the event multiplicity and is about 5--6.5$\%$ for minimum-ionising particles emerging from the interaction point and reaching the outer radius of the TPC~\cite{ALICEperformance}. For charged-particle tracks reconstructed from their hits in the TPC and ITS, the transverse-momentum (\pt) resolution ranges from about 1$\%$ at $\pt=1$ GeV/$c$ to about 2$\%$ at 10 GeV/$c$~\cite{ptSpectraRAA}. The resolution in the measurement of the distance of closest approach (DCA) of primary tracks to the primary collision vertex, projected on the transverse plane, ranges from about 200 $\rm \mu$m at $\pt=0.2$ GeV/$c$ to about 10 $\rm \mu$m at 10 GeV/$c$~\cite{ALICEperformance}.

The PID is complemented by the Time-Of-Flight (TOF) system~\cite{TOF}. This detector is made of Multi-gap Resistive Plate Chambers and is located at a radial distance of 3.7 m from the nominal interaction point. The TOF detector measures the arrival time of particles relative to the event collision time provided by the TOF detector itself or by the T0 detectors, two arrays of Cherenkov counters located at forward and backward rapidities~\cite{CollisionTime}. The TOF detector is used in combination with the ITS for pile-up rejection, mostly from collisions which belong to different bunch crossings (out-of-bunch), by requiring that at least one of the strange hadron decay particles has a reconstructed track with an associated hit in the TOF detector, as described in Sec.~\ref{subsec:CascadeAndV0Selection}.

Collision events are triggered by two plastic scintillator arrays, V0A and V0C~\cite{VZEROPerformance}, located on both sides of the interaction point, covering the pseudorapidity regions \mbox{2.8 $< \eta <$ 5.1} and $-3.7 <  \eta < -1.7$, respectively. Each array consists of four concentric rings, each ring comprising eight cells with the same azimuthal coverage. The V0A and V0C scintillators can be used to characterise events on collision multiplicity ~\cite{CentralityDetermination}.
Given the $\eta$-gap with respect to the midrapidity region, such an estimator can be effectively used to select events with different activities in the forward region even when the multiplicity at midrapidity is fixed.

The effective energy is accessible by measuring the energy carried by nucleons emitted at forward rapidities using two zero-degree calorimeters (ZDC)~\cite{CERN-LHCC-99-005,ALICEperformance}.
These identical detectors, placed at $\pm$112.5 m from the ALICE interaction point, on both sides, consist of a neutron (ZDC-N) and a proton (ZDC-P) calorimeter.
The ZDC-N calorimeters cover the pseudorapidity range $|\eta|>8.8$, while the geometrical coverage of the ZDC-P calorimeters is $6.5<|\eta|<7.4$.
In this work, the effective energy is estimated using only the neutron calorimeter, as will be discussed in the following sections.

%% file: Chapters/analysis.tex
\section{Data sample}
\label{sec:analysis}
The data used for this analysis were collected in 2015, 2017, and 2018 during the LHC pp runs at ${\sqrt{s} = 13 ~\rm TeV}$, in specific data-taking periods where the ZDC detectors were switched on.
A limited half-crossing angle of the beams in the vertical plane was applied in these runs, corresponding to $+45~\mu$rad for 2015 data and $+70~\mu$rad for 2017 and 2018 data.
This configuration guarantees that all the neutrons emitted at very forward rapidities fall within the ZDC-N geometric acceptance.
The acceptance of the neutron calorimeter is not affected provided that the vertical half-crossing angle is smaller than $+60~\mu$rad for a nominal vertex vertical position on the LHC axis ($y_{\textrm{vtx}}=0 ~\rm mm$), and smaller than $+79~\mu$rad for a position of $y_{\textrm{vtx}} = - 1~ \rm mm$.
The $y_{\textrm{vtx}}$ was equal to $0~\rm mm$ for 2015 data, and equal to $-1~\rm mm$ for 2017 and 2018 data.
A minimum bias (MB) event trigger was used, which requires coincident signals in the V0 detectors to be synchronous with the bunch crossing time defined by the LHC clock.
In order to ensure full geometrical acceptance of central barrel detectors and reject background collisions, the coordinate of the primary vertex along the beam axis is required to be within 10 cm from the nominal interaction point. The contamination from beam-induced background is removed during the offline analysis using the timing information from the V0 detectors and taking into account the correlation between the number of tracklets, short track segments reconstructed at midrapidity, and the number of hits in the SPD detector~\cite{ALICEperformance}. The data periods used in the analysis are characterised by a
$\mu$ value (average number of proton--proton interactions per bunch crossing) which ranges from $\sim1\%$ to a maximum value of $\sim14\%$. Events with more than one reconstructed primary interaction vertex in the same bunch crossing (in-bunch pile-up) identified from tracklets in the SPD are tagged as pile-up and removed from the analysis~\cite{ALICEperformance}. In addition, events with pile-up occurring during the drift time of the TPC are rejected based on the correlation between the number of SDD and SSD clusters and the total number of clusters in the TPC, as described in Ref.~\cite{JPsiFlow}. To further suppress the pile-up contribution, mostly from out-of-bunch collisions, it is requested that at least one of the tracks from the decay products of the (multi-)strange hadron under study is matched in either the ITS or the TOF detector.
The results are reported for the INEL$>$0 event class, defined by requiring at least one charged particle within the pseudorapidity interval $|\eta|<$1, corresponding to $\sim$75$\%$ of the total inelastic cross section. A total number of 1.29$\times10^{8}$ MB events were selected after applying these requirements.

\section{Event classification}
\label{subsec:AnalysisStrategy}
The measurement of strange hadron production presented in this paper is performed as a function of the charged-particle multiplicity density at midrapidity and of the leading ("zero degree") energy.
This approach is aimed at investigating the connection of strangeness production to global properties of the pp collision, experimentally measured through the forward energy in the event, and to local effects, characterised by the charged-particle multiplicity at midrapidity.
The dynamical range between measured global and local event properties spans a pseudorapidity gap of 8 units, equivalent to the separation between the ALICE ZDC and the central barrel.

Through a two-dimensional analysis in terms of these observables, this approach considers events with defined global properties, grouping collisions in classes with similar average values and profile distributions of the zero-degree energy, and explores different local conditions in terms of charged-particle multiplicity at midrapidity: in this case, strangeness production can be studied in relation to jet production and local fluctuations in the hadronisation process.
In a complementary way, this work considers events with defined local properties, i.e.~similar multiplicity of particles produced at midrapidity, and explores different forward energy deposits: in this case, the local environment in which strangeness is produced is fixed in terms of parton density, while global properties of the event may vary, for example, due to the available energy in the collision or the number of parton--parton interactions that occurred.

\begin{figure}[htbp]
    \centering
    \includegraphics[width = 0.9\textwidth]{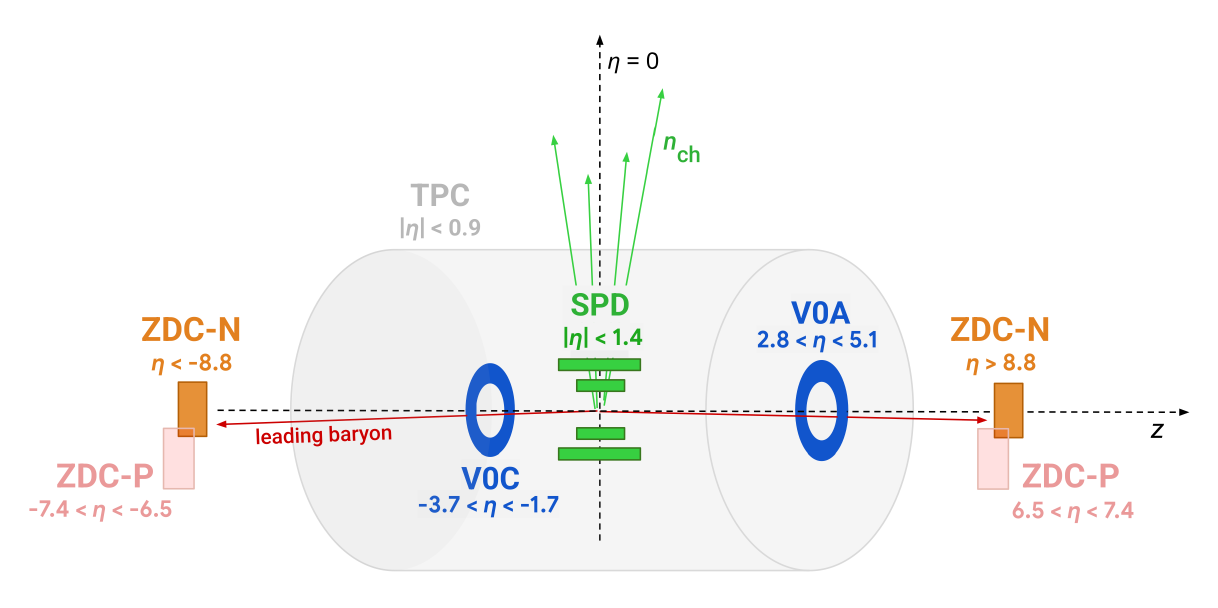}
    \caption{Illustration of the relative positions and pseudorapidity coverage of the SPD, TPC, ZDC-N, ZDC-P and V0 detectors used in this study. Dimensions are not to scale. }
    \label{fig:sketch}
\end{figure}

The collision events characterised by defined local-multiplicity and forward energy properties are selected using an approach which resembles an event-shape engineering technique~\cite{PhysRevC.93.034916}.
While the final results will be presented as a function of \dndeta and of the zero-degree energy, the sample of pp collisions is divided into classes defined through two independent estimators: one covering the forward rapidity region (V0M) and one exploiting information at midrapidity (SPDClusters).
The V0M estimator is based on the signal amplitude measured by the V0 detectors (V0A and V0C), which reflects the total charge deposited in the forward region.
The SPDClusters estimator is based on the number of hits (clusters) measured at midrapidity by the two SPD layers.
The V0M detectors are positioned closer to midrapidity compared to the ZDC, resulting in a smaller $\eta$ gap between the SPD and V0M than the 8-unit separation between the forward calorimeters and the central barrel.
Based on the signals measured by the V0 and SPD detectors, the events are divided into percentile classes, which reflect the fraction of events in each interval over the total number of events.
Figure~\ref{fig:sketch} illustrates the relative position and pseudorapidity coverage of the detectors used to engineer event classes in this work.

To check if the selections based on the V0M and SPDClusters estimators introduced any biases which could alter the relative abundances of different hadron species the evolution of charged- and neutral-kaon abundances with multiplicity was checked using a PYTHIA 8 Monte Carlo sample which includes the simulation of the detector response.
The number of reconstructed charged and neutral kaons were found to be very similar, independently of the multiplicity, as expected due to their similar masses.

The charged-particle multiplicity density is measured from SPD tracklets in the pseudorapidity interval ${|\eta| < 0.5}$ using the technique described in Ref.~\cite{pseudorapidityDensity}. Note that in the following, for simplicity, the average charged pseudorapidity density $\langle \dndeta \rangle$ is reported as $\langle n_{\rm ch} \rangle$. The very-forward energy ($ZN$) is measured as the amplitude of the signal detected only by neutron calorimeters, to avoid the strong acceptance limitations related to the beam optics deflection of charged particles, which can lead to underestimating the energy event-by-event in the ZDC-P calorimeter.
As a result, the correlation of ZDC-P energy and the energy of particles emitted in the forward direction would be affected from larger systematic uncertainties than the ZDC-N case.
Considering acceptance limitations, the ZDC signal in pp collisions is not very effective to classify events in percentile selections on an event-by-event basis.
On the contrary, it works very well on average to characterise global properties of the collisions, when event classes are built with independent estimators.
In Pb--Pb collisions, the energy calibrations of ZDC-N spectra are performed using the narrow peaks measured from the detection of single neutrons. Instead, in pp collisions there is no reliable way to calibrate the calorimeter spectra in energy units without introducing model dependencies and large uncertainties.
In this paper, self-normalised quantities are used, namely signals normalised to their average minimum-bias value, which allow one to overcome this problem and to obtain results which are directly comparable to model predictions.

\begin{figure}[htbp]
    \centering
    \subfigure[]{\includegraphics[width = 0.98\textwidth]{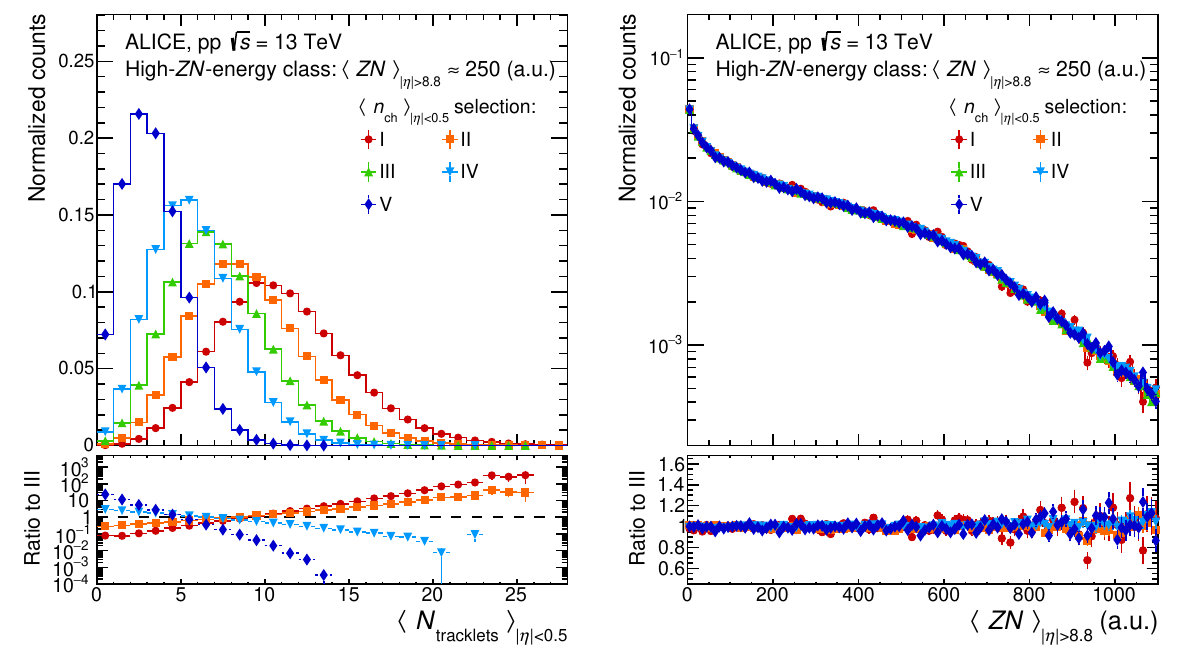}}
    \subfigure[]{\includegraphics[width = 0.98\textwidth]{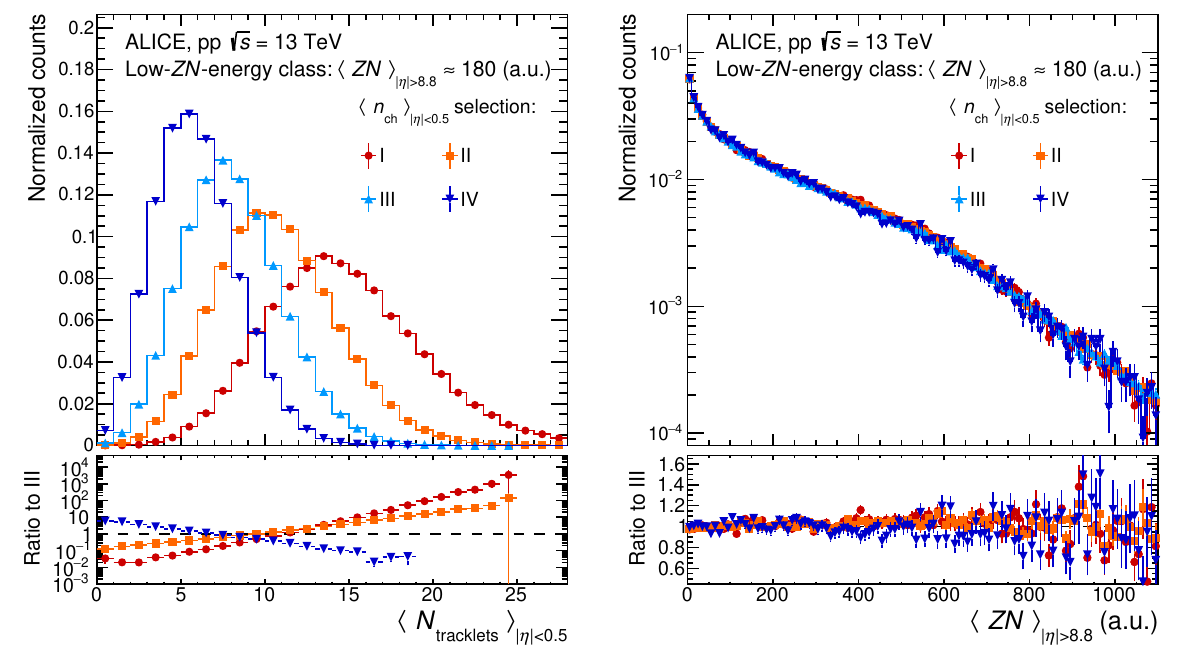}}
    \caption{Distributions of raw tracklets and $ZN$ energy for high- (a) and low- (b) $ZN$-energy event classes, which are based on different combinations of selections on the V0M and SPDClusters percentiles (see text for details).}
    \label{fig:distributionszn}
\end{figure}

\begin{figure}[htbp]
    \centering
    \subfigure[]{\includegraphics[width = 0.98\textwidth]{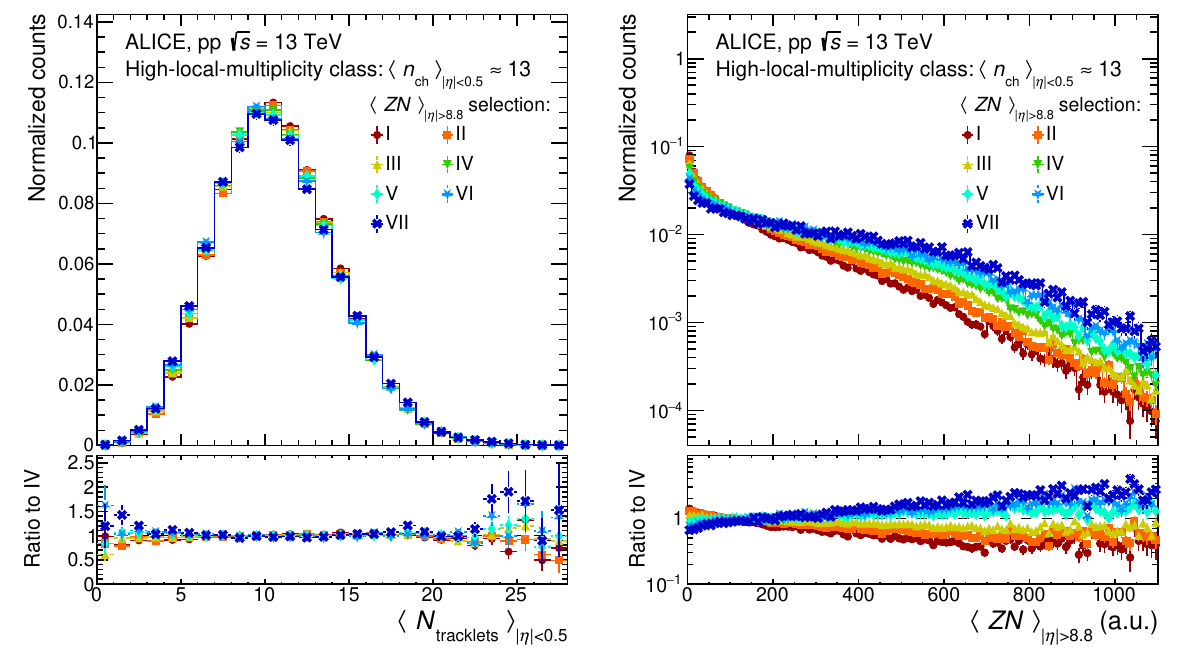}}
    \subfigure[]{\includegraphics[width = 0.98\textwidth]{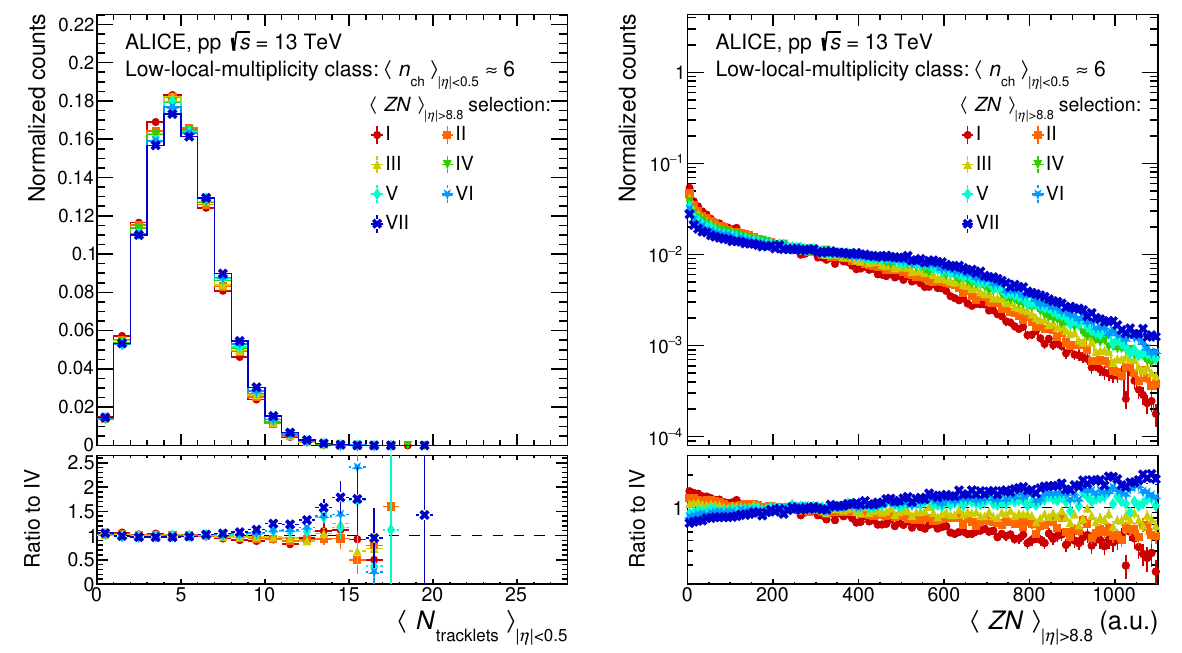}}
    \caption{Distributions of raw tracklets and $ZN$ energy for high- (a) and low- (b) local-multiplicity event classes, which are based on different combinations of selections on the V0M and SPDClusters percentiles (see text for details).}
    \label{fig:distributionsnch}
\end{figure}

Five different types of classifications are defined for the analysis, based on different combinations of selections on the V0M and SPDClusters percentiles, in order to allow the study of strange hadron production as a function of particle density at midrapidity and leading energy. The first type of classification is performed using V0M percentile intervals and results into event classes with increasing average local multiplicity and by decreasing average very-forward energy deposits. The event classes are labelled I, II, III, etc., where class I corresponds to the highest multiplicity and lowest $ZN$ energy class. These selections will be referred to as "standalone".
The second and third types of classification are performed by selecting events with similar average values of energy detected in the ZDC-N calorimeters, but different $\langle n_{\rm ch} \rangle$. This is achieved using a combination of the V0M and SPDClusters estimators, and, also in this case, class I corresponds to the highest multiplicity class.
In each event class, the distributions of the $ZN$ energy and of the number of raw tracklets reconstructed at midrapidity were studied. The selections were engineered to ensure not only a consistent average value of $ZN$ energy in the different classes, but also similar distributions.
Two classifications were built, one consisting of events with high ZN energy and one of events with low ZN energy. They will be referred to as "high-/low-$ZN$-energy" selections".
The distributions of the number of tracklets and $ZN$ energy for these classifications are reported in Fig.~\ref{fig:distributionszn}.
For the fourth and fifth types of classification, the sample is divided into classes characterised by similar average values of charged-particle pseudorapidity density at midrapidity and different $ZN$ energies, where class I corresponds to the lowest $ZN$ energy.
In this case, the selections are defined by fixing the SPD clusters in a narrow percentile range and varying the V0M estimator. Also in this case, the selections are defined so that the different classes have not only a consistent average value of local multiplicity, but also similar distributions.
These classifications will be referred to as "high-/low-local-multiplicity" selections, and the distributions of the number of tracklets and $ZN$ energy in these classes are reported in Fig.~\ref{fig:distributionsnch}.
In this way, a total of five types of event classifications are defined, summarised in Table~\ref{Table:eventclasses}.


\begin{table}[htbp]
\footnotesize
\begin{center}
\centering
\renewcommand{\arraystretch}{1.4}
\caption{Types of classifications defined according to different combinations of V0M and SPDClusters estimators (see text for details). For each event class, the third and forth columns summarise the fraction of the $\rm INEL>0$ cross section corresponding to the V0M and SPDClusters estimators, respectively. The average values of  $\langle n_{\rm ch} \rangle$ and $ZN$ energy calculated for each event class are reported in the fifth and sixth columns, respectively.}
\begin{tabular}{| c | c | c | c | c | c | c |}
\hline
 & & $\sigma / \sigma_{\rm INEL>0}$ &  $\sigma / \sigma_{\rm INEL>0}$  &   & \\
Classification type & Event Class &  V0M selection & SPDClusters selection  & $\langle n_{\rm ch} \rangle_{|\eta|<0.5}$ &  $\langle ZN \rangle_{|\eta|>8.8}$ (a.u.)  \\
\hline
\hline
  INEL$>$0  &  & 0--100.0$\%$ &  0--100.0$\%$   &  6.89$\pm$0.11 &     280$\pm$8   \\
\hline
\hline
Standalone  & I & 0--0.90\% & 0--100.0\% & 25.75$\pm$0.40 & 80$\pm$2 \\
\hline
 & II & 0.90--4.5\% & 0--100.0\% & 19.83$\pm$0.30 & 106$\pm$3 \\
\hline
 & III & 4.5--8.9\% & 0--100.0\% & 16.12$\pm$0.24 & 136$\pm$4 \\
\hline
 & IV & 8.9--13.4\% & 0--100.0\% & 13.76$\pm$0.21 & 163$\pm$5 \\
\hline
 & V & 13.4--17.9\% & 0--100.0\% & 12.06$\pm$0.18 & 186$\pm$6 \\
\hline
 & VI & 17.9--26.8\% & 0--100.0\% & 10.11$\pm$0.15 & 217$\pm$7 \\
\hline
 & VII & 26.8--35.8\% & 0--100.0\% & 8.07$\pm$0.12 & 254$\pm$8 \\
\hline
 & VIII & 35.8--44.8\% & 0--100.0\% & 6.48$\pm$0.09 & 287$\pm$9 \\
\hline
 & IX & 44.8--63.5\% & 0--100.0\% & 4.64$\pm$0.06 & 327$\pm$10 \\
\hline
 & X & 63.5--100.0\% & 0--100.0\% & 2.52$\pm$0.03
 & 369$\pm$11 \\
\hline
\hline
 High-$ZN$-energy & I & 35.8--54.0\% & 0--17.9\% & 13.92$\pm$0.34 & 256$\pm$8 \\
\hline
 & II & 26.8--63.5\% & 8.9--26.8\% & 11.29$\pm$0.27 & 251$\pm$8 \\
\hline
 & III & 26.8--44.8\% & 17.9--35.7\% & 9.05$\pm$0.22 & 254$\pm$8 \\
\hline
 & IV & 17.9--44.8\% & 26.8--44.7\% & 7.27$\pm$0.17 & 256$\pm$8 \\
\hline
 & V & 0--26.8\% & 44.7--100.0\% & 4.28$\pm$0.10 & 255$\pm$8 \\
 \hline
 \hline
  Low-$ZN$-energy & I & 17.9--26.8\% & 0--8.9\% & 18.73$\pm$0.43 & 180$\pm$5 \\
\hline
 & II & 8.9--26.8\% & 8.9--17.9\% & 13.6$\pm$0.31 & 179$\pm$5 \\
\hline
 & III & 0--17.9\% & 17.9--26.8\% & 10.43$\pm$0.23 & 175$\pm$5 \\
\hline
 & IV & 0--8.9\% & 26.8--44.7\% & 7.74$\pm$0.17 & 173$\pm$5 \\
\hline
\hline
High-local-multiplicity  & I & 0--4.5\% & 8.9--17.9\% & 13.97$\pm$0.16 & 121$\pm$4 \\
\hline
 & II & 4.5--8.9\% & 8.9--17.9\% & 13.79$\pm$0.17 & 141$\pm$4 \\
\hline
 & III & 8.9--17.9\% & 8.9--17.9\% & 13.65$\pm$0.17 & 167$\pm$5 \\
\hline
 & IV & 17.9--26.8\% & 8.9--17.9\% & 13.48$\pm$0.17 & 197$\pm$6 \\
\hline
 & V & 26.8--35.8\% & 8.9--17.9\% & 13.35$\pm$0.17 & 224$\pm$7 \\
\hline
 & VI & 35.8--44.8\% & 8.9--17.9\% & 13.24$\pm$0.17 & 251$\pm$8 \\
\hline
 & VII & 44.8--100.0\% & 8.9--17.9\% & 13.15$\pm$0.16 & 286$\pm$9 \\
 \hline
 \hline
Low-local-multiplicity & I & 0--17.9\% &  35.7--44.7\% & 6.19$\pm$0.07 & 210$\pm$6 \\
\hline
 & II & 17.9--26.8\% &  35.7--44.7\% & 6.15$\pm$0.07 & 239$\pm$7 \\
\hline
 & III & 26.8--35.8\% &  35.7--44.7\% & 6.14$\pm$0.07 & 263$\pm$8 \\
\hline
 & IV & 35.8--44.8\% &  35.7--44.7\% & 6.13$\pm$0.08 & 285$\pm$9 \\
\hline
 & V & 44.8--54.0\% &  35.7--44.7\% & 6.09$\pm$0.08 & 306$\pm$9\\
\hline
 & VI & 54.0--63.5\% &  35.7--44.7\% & 6.07$\pm$0.09 & 325$\pm$9 \\
\hline
 & VII & 63.5--100.0\% &  35.7--44.7\% & 6.07$\pm$0.09 & 352$\pm$11 \\
\hline
\end{tabular}
\label{Table:eventclasses}
\end{center}
\normalsize
\end{table}

\newpage
\section{Phenomenological models and comparison to the data}
\label{subsec:Phenomodels}

Several phenomenological models have been investigated to understand the underlying mechanisms of strangeness production in hadronic collisions. Some of these models adapt concepts from heavy-ion physics, such as statistical hadronisation and hydrodynamic expansion (including core--corona models), to smaller systems. Alternatively, other models extend proton--proton collision descriptions, based on hard-partonic interactions, underlying event, and string fragmentation, to higher multiplicity regimes, introducing new production mechanisms.
In this context, PYTHIA~\cite{pythia8generic} is a general purpose Monte Carlo event generator which implements colour string fragmentation at its core. In the Monash tune of PYTHIA 8, MPIs have been introduced to describe charged-particle pseudorapidity densities in high energy hadronic collisions, and colour reconnection~\cite{PhysRevD.36.2019} mechanisms have been implemented to account for \pt~spectra modifications in events featuring a larger or smaller final state multiplicity.
LHC results triggered several new paths of phenomenological investigation, targeted at reproducing the multiplicity dependence observed for strangeness and baryon production yields. For this purpose, sophisticated colour reconnection mechanisms allowing for three-leg junctions~\cite{ColRecBeyondLead} were introduced to better describe baryon production yields. In addition, a modified string tension in dense QCD environments~\cite{bierlich2015effects} has been proposed to describe strangeness enhancement in hadronic collisions, allowing for overlapping strings to interact forming colour ropes. In the following, this upgraded version of PYTHIA 8 will be referred to as QCD-CR+Ropes. It is worth noting that when switching-on the rope mechanism, improved CR has to be enabled. Therefore, the features distinguishing the Monash and QCD-CR+Ropes tunes may come from one of these additional mechanisms, or from their interplay.

The comparison of phenomenological models to data is performed at the generator level, i.e.~using only the kinematic information of the generated particles, without simulating their passage through the ALICE detector and reconstructing each event as for real collisions.
The generated samples consist of $2\times10^{9}$ pp collision events at $\sqrt{s} = 13 \rm ~TeV$ simulated with the Monash tune of the PYTHIA 8 generator and $2\times10^{9}$ simulated with PYTHIA 8 QCD-CR+Ropes.
The analysis procedure reproduces the one used for the data, starting from the generator-level information.
The leading energy is simulated considering the acceptance of the ZDC-N calorimeter, i.e.~as the sum of the energy of neutral primary particles in the pseudorapidity region $|\eta| > 8.8$.
The V0M signal is simulated by counting primary charged particles generated in the detector acceptance.
The SPDClusters estimator is simulated considering both primary and secondary particles from weak decays produced in the acceptance of the SPD layers.
Using a Monte Carlo sample which includes the simulation of the detector response, the agreement between simulated quantities (percentile estimators and $ZN$ energy) and reconstructed ones was then checked. The observed differences are small, with a standard deviation of at most 15$\%$.

\subsection{Forward energy as a function of charged-particle production at midrapidity}
\label{subsec:ResultsNchZDC}
\begin{figure}[!hbt]
    \centering
    \includegraphics[width = 0.62\textwidth]{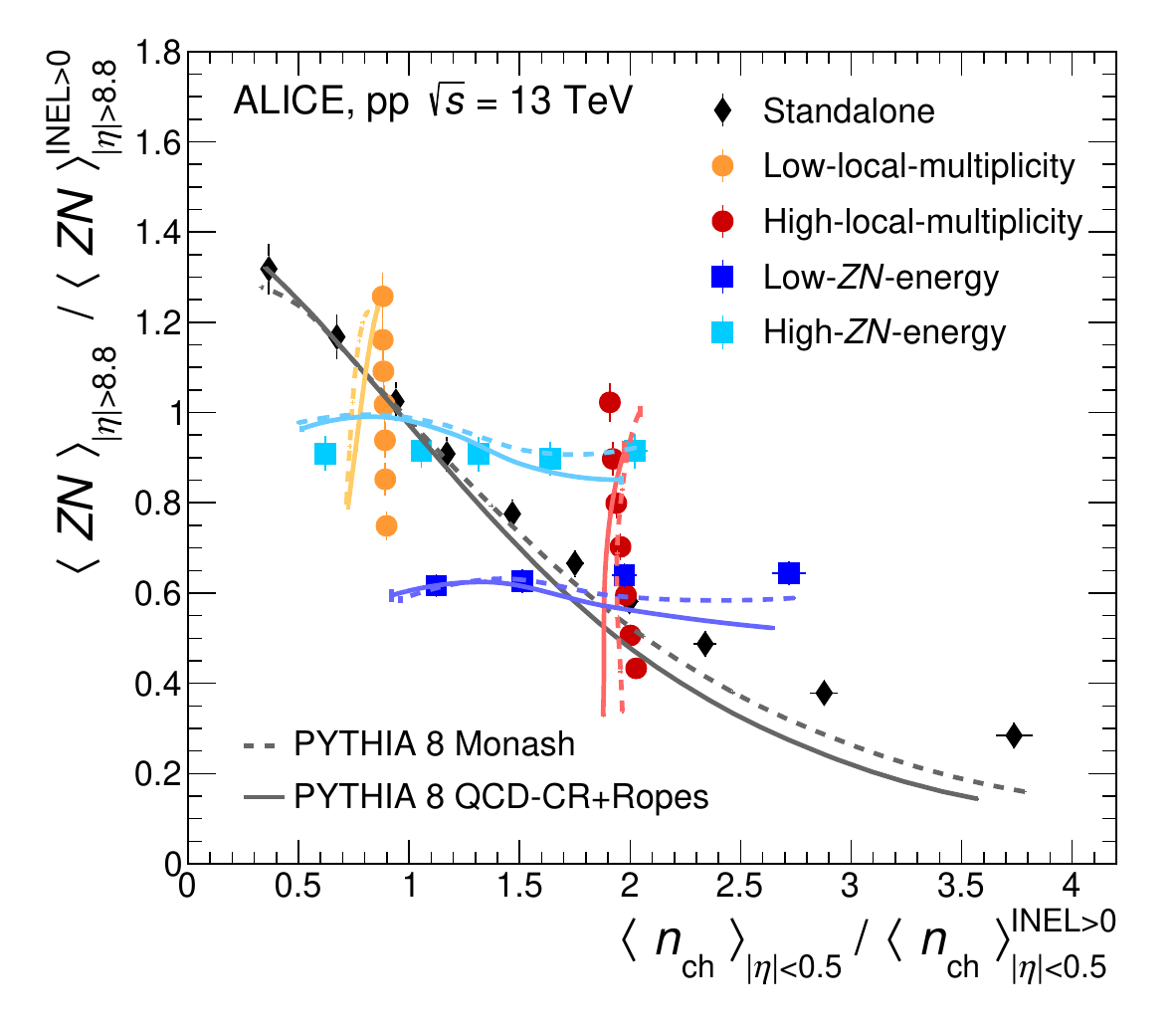}
    \caption{Self-normalised $\langle ZN \rangle$ signal as a function of the self-normalised charged-particle multiplicity measured at midrapidity $\langle n_{\rm ch} \rangle$. Statistical errors are negligible, only systematic uncertainties are shown.}
    \label{fig:znvsnch}
\end{figure}

Figure~\ref{fig:znvsnch} summarises the relation between the self-normalised $\langle ZN \rangle$ signal and $\langle n_{\rm ch} \rangle$ for the event classes listed in Table~\ref{Table:eventclasses}.
The results obtained in this work with the standalone selections are in agreement with ALICE results reported in Ref.~\cite{correlationZDCdNdeta}, where the forward energy detected by the ZDC was studied as a function of the charged-particle multiplicity produced at midrapidity in pp and p--Pb collisions~\cite{correlationZDCdNdeta}.
The two quantities are anti-correlated: the higher the activity measured at midrapidity, the smaller the forward energy.
This observation can be interpreted as a positive correlation of charged-particle production with the effective energy, assuming the deposit in the ZDC calorimeters is proportional to the energy of leading particles.
As discussed in Sec.~\ref{subsec:AnalysisStrategy}, in this work multi-differential event classes are defined in order to disentangle the correlation between midrapidity particle production and the energy measured at forward rapidity.
In the high-local-multiplicity classes, events with large $\langle n_{\rm ch} \rangle$ show $ZN$ energy values which cover a range between about 0.4 and 1.0 times the minimum-bias value, while for the low-local-multiplicity classification, the self-normalised $ZN$ values range between about 0.7 and 1.3.
The total forward energy interval covered by the standalone selections ranges between about 0.3 and 1.3 times the minimum-bias value. Therefore, also at fixed multiplicity, this analysis is able to select events covering a significant portion of the $ZN$ signal interval that is covered by the standalone selections.
Similarly, the self-normalised multiplicity values in the classes selected requiring high $ZN$ energy cover a range between about 0.5 and 2.0, and between about 1.1 and 2.7 for the low $ZN$.
The total average multiplicity interval covered by the standalone selection ranges between 0.4 and 3.8 times the minimum-bias value, therefore, also at fixed $ZN$ energy, a significant range of accessible multiplicities is covered.
Data are compared with MC simulations based on the PYTHIA 8 event generator with the Monash and QCD-CR+Ropes tunes.
Both tunes are able to describe the overall decreasing trend of the standalone selections.
In particular, a good agreement between data and MC is observed at low $n_{\rm ch}$ corresponding to high $ZN$ energies, while at low $ZN$ energies the agreement is worse, with both PYTHIA 8 tunes underestimating the forward energy by up to 40$\%$.
It is worth noting that the two tunes can qualitatively describe the behaviour of the data points in all the multi-differential selections.
The high-/low-local-multiplicity classes select simulated collisions with fixed multiplicity, underestimating by up to 10$\%$ the multiplicity values with respect to the data points and covering a $ZN$ energy range similar to the measurements.
The high-$ZN$-energy and low-$ZN$-energy event classes select simulated collisions with forward energy values within $\pm 10\%$ with respect to the measured ones, covering a range of midrapidity multiplicity similar to the data points.
In general, the agreement between data and MC is found to be slightly better for PYTHIA 8 with Monash tune, however, one can observe that including the hadronisation mechanisms implemented in the QCD-CR+Ropes tune does not have a big impact on the description of the correlation between the multiplicity and the leading energy.

\subsection{Sensitivity to global and local effects in PYTHIA 8}
\label{subsec:LEvsMPI}
\begin{figure}[!ht]
    \centering
    \includegraphics[width = 0.95\textwidth]{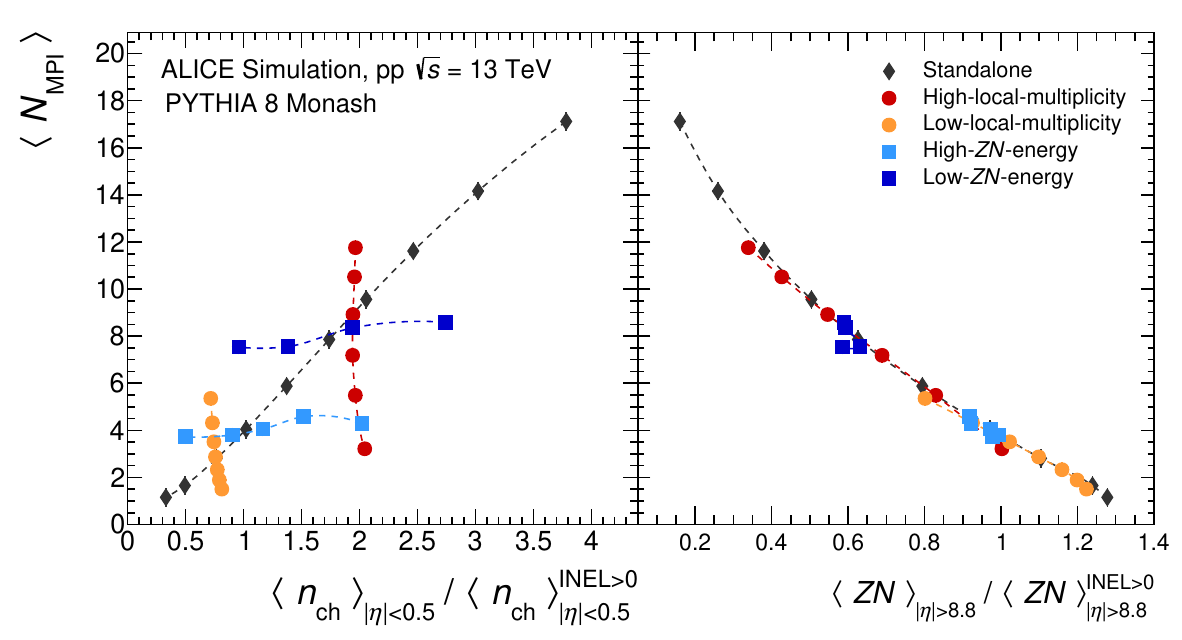}
    \caption{Average number of MPIs from PYTHIA 8 simulations as a function of the self-normalised midrapidity multiplicity $\langle n_{\rm ch} \rangle$ (left) and of the self-normalised forward energy $\langle ZN \rangle$ (right) in the different multi-differential classes introduced in this work. Statistical uncertainties are shown as vertical bars, but they are hardly visible in the figures.}
    \label{fig:MPI}
\end{figure}

According to the PYTHIA 8 event generator, pp collision events with high multiplicity at midrapidity mainly originate either from multiple semi-hard MPIs occurring within the same pp collision or from multi-jet final states (hard processes).
In particular, the presence of jets at midrapidity in the final state can be studied in the model by considering, as an example, the average transverse momentum of charged pions $(\langle p_{\rm T}^{\pi} \rangle_{|y|<0.5})$, proxy for the \pt of the hard parton-scattering process.
To investigate the sensitivity of the applied selections to global and local effects, the correlation of the number of MPIs and of the average \pt of pions with the local charged-particle multiplicity and the leading energy is studied using the multi-differential approach introduced in this paper.
The results presented in this section are obtained with the PYTHIA 8 event generator with the Monash tune, but qualitatively similar results are produced with the QCD-CR+Ropes tune.
The correlation between the effective energy and the number of MPIs can provide important insights into the interpretation of the results on strange hadron production discussed in this paper.
In fact, the string hadronisation processes implemented in PYTHIA are strongly influenced by the number of MPIs occurred in the collision and the average energy produced at very forward rapidity was found to decrease with an increasing number of MPIs~\cite{correlationZDCdNdeta}.
Figure~\ref{fig:MPI} shows the average number of parton--parton interactions as a function of the midrapidity multiplicity (left) and of the forward energy in the ZDC-N acceptance (right) in the event classes introduced in this work.
In the high-/low-local-multiplicity event classes (orange and red circles), the average number of MPIs increases at fixed midrapidity multiplicity with decreasing leading energies.
On the other hand, once events with defined leading energy are selected (azure and blue squares), the average number of MPIs does not change significantly with the midrapidity multiplicity, exhibiting a rather flat trend in the left panel of Fig.~\ref{fig:MPI}.
It is worth noting that the average number of MPIs shows a universal dependence with the leading energy, i.e.~common for all differential selections, as it can be seen in the right panel of Fig.~\ref{fig:MPI}.
These results show that the leading energy is a powerful observable to probe the dependence of particle production on the number of MPIs in PYTHIA.

\begin{figure}[!ht]
    \centering
    \includegraphics[width = 0.95\textwidth]{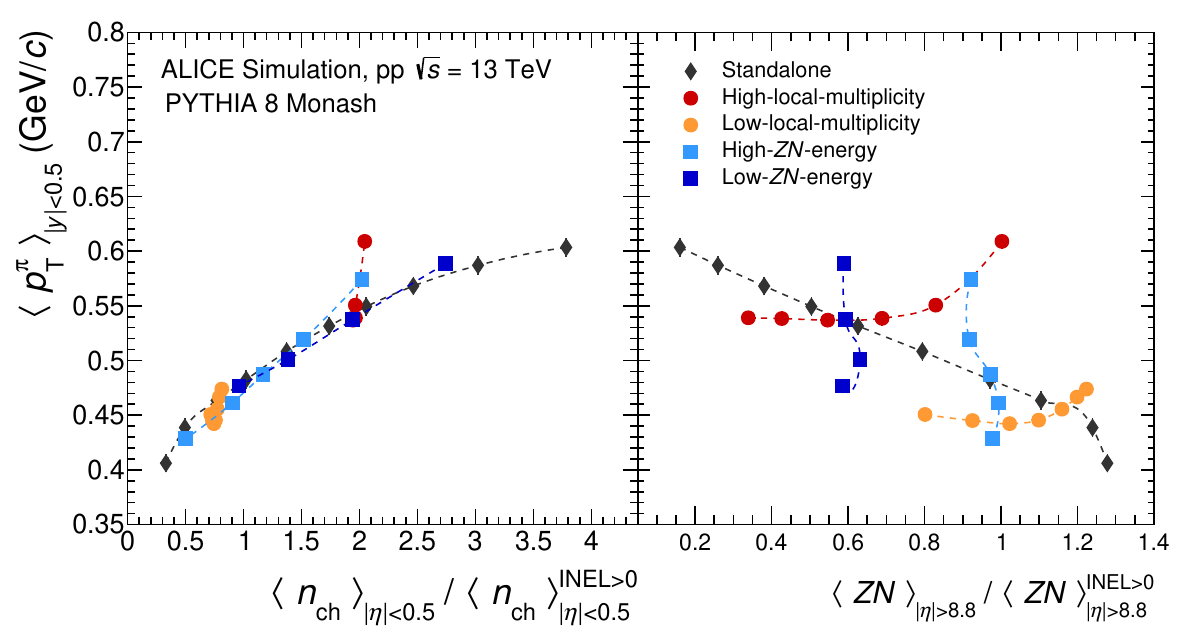}
    \caption{Average transverse momentum of charged pions from PYTHIA 8 simulations as a function of self-normalised midrapidity multiplicity $\langle n_{\rm ch} \rangle$ (left) and of the self-normalised forward energy $\langle ZN \rangle$ (right) in the different multi-differential classes introduced in this work. Statistical uncertainties are shown as vertical bars, but they are hardly visible in the figures.}
    \label{fig:meanpt}
\end{figure}

Figure~\ref{fig:meanpt} shows the average transverse momentum of charged pions as a function of the charged-particle multiplicity at midrapidity (left) and of the leading energy in the ZDC-N acceptance (right).
Once events with defined leading energy are selected (high-/low-$ZN$-energy classes), the average \pt is found to increase with the midrapidity multiplicity, following a common trend for all event classes.
On the other hand, for similar midrapidity multiplicities, the $(\langle p_{\rm T}^{\pi} \rangle_{|y|<0.5})$ shows only a very mild dependence with the leading energy.
This observation suggests that local phenomena, such as jets at midrapidity, are correlated with local observables, such as the charged-particle multiplicity, and rather independent from global properties of the event, e.g. the available energy in the collision.
In summary, the studies on the output of the PYTHIA 8 event generator indicate that a two-dimensional analysis as a function of the midrapidity multiplicity and the very-forward energy can disentangle the connection of a given phenomenon to global and local properties of the collision.

\section{Cascade and \texorpdfstring{V$^{0}$}{} selection}
\label{subsec:CascadeAndV0Selection}
The \pt spectra of $\rm K_{S}^0$, $\Lambda$($\overline{\Lambda}$), and $\Xi^{-}$($\overline{\Xi}^{+}$) are measured reconstructing the (multi-)strange hadrons at midrapidity ($|y|$ $<$ 0.5) using their weak-decay channels:

\begin{list}{$\bullet$}{}
\item $\rm K_{S}^0 \rightarrow \pi^{+}+\pi^{-}$  \ \ \ Branching Ratio  $=$ $(69.20 \pm 0.05) \%$,
\item $\Lambda \rightarrow \mathrm{p} + \pi^{-}$  \ \ \ \ Branching Ratio  $=$ $(63.9 \pm 0.5) \%$,
\item $\Xi^{-} \rightarrow \Lambda + \pi^{-}$             \ \ \ Branching Ratio  $=$ $(99.887 \pm 0.035)\%$.
\end{list}
along with their charge conjugates~\cite{ParticleDataGroup:2024cfk}.
The $\Lambda$($\overline{\Lambda}$) and $\rm K_S^0$ candidates are reconstructed using the standard ALICE weak-decay finder. This algorithm searches for neutral weak-decay topologies, called $\mathrm{V}^0$, by reconstructing oppositely-charged particle tracks originating from a displaced vertex, as described in Refs.~\cite{strangenessEnhancementNature,strangenessEnhancementpp}. In the case of the $\Xi^{\pm}$ baryons, the cascade finder is used, which searches for a pair composed of one reconstructed $\mathrm{V}^0$ and one additional charged particle (bachelor), pointing inward to the same displaced vertex. In the case of a $\mathrm{V}^0$ decay vertex located inside the ITS volume, at least one hit in any of the ITS layers is requested in the reconstruction of the charged tracks originating from the $\mathrm{V}^0$ decay.
The reconstructed tracks, selected in the pseudorapidity region $|\eta| < 0.8$, are required to fulfil a set of quality criteria, such as to produce signal in at least 70 TPC readout pads out of a maximum of 159. Moreover, the fraction of TPC pad rows with a signal over the number of clusters expected based on the reconstructed trajectory is required to be at least 80$\%$. This condition ensures that tracks do not have large gaps in the associated hits in the radial direction.

To reduce the combinatorial background, a set of topological selections is applied to the reconstructed hadrons. The analysis relies on the same set of standard cuts used in Ref.~\cite{Acharya_2020}. For the measurement of $\mathrm{V}^0$s, the distance of closest approach between the $\mathrm{V}^0$ daughter tracks is required to be less than 1 standard deviation, the DCA between each $\mathrm{V}^0$ daughter and the primary collision vertex larger than 0.06~cm and the radial distance between primary and secondary vertices larger than 0.5~cm. The cosine of the pointing angle, defined as the angle between the vector connecting the primary and secondary vertices and the total $\mathrm{V}^0$ reconstructed momentum, is required to be larger than $0.995$. A proper-lifetime selection is applied to the reconstructed $\Lambda$ ($\rm K_S^0$) candidates by requiring $mL/p$ to be lower than 20 (30)~cm, where $m$ is the candidate mass, $L$ is the linear distance between the candidate decay point and the primary vertex, $p$ the total momentum.
Candidates compatible with the alternative $\mathrm{V}^0$ hypothesis are rejected if they lie within $\pm$5 MeV/$c^{2}$ ($\pm$10 MeV/$c^{2}$) of the nominal $\Lambda$ ($\rm K_S^0$) mass. For cascades, the DCA between the bachelor track and the primary vertex is required to be larger than 0.04 cm, the DCA between the $\mathrm{V}^0$ and the primary vertex larger than 0.06 cm, while the DCA between the bachelor track and $\mathrm{V}^0$ lower than 1.3~cm. The DCA between the $\mathrm{V}^0$ daughter meson (baryon) and the primary collision vertex is required to be larger than 0.04 (0.03)~cm, while the DCA between the $\mathrm{V}^0$ daughter tracks is required to be less than 1.5 standard deviations. In addition, a minimum radial distance of 0.6 (1.2)~cm between the primary and cascade ($\mathrm{V}^0$) weak-decay vertex is required. The cosine of the pointing angles of both cascade and $\mathrm{V}^0$ is required to be larger than 0.97. The $\mathrm{V}^0$ produced in cascade decays are required to match an invariant mass window of $\pm$ 8 MeV/$c^{2}$ with respect to the nominal $\Lambda$ mass, and a proper-lifetime selection is applied, requiring $mL/p$ lower than $3c\tau$, where $\tau$ is the mean lifetime of the particle. In addition, in order to reject the residual out-of-bunch pile-up background on the measured yields, it is requested that at least one of the tracks from the decay products of the (multi-)strange hadron under study is matched in either the ITS or the TOF detector. The selection criteria applied for this measurement are optimised based on detailed studies done on Monte Carlo simulations and are similar to those already used in previous measurements~\cite{strangenessEnhancementNature,strangenessEnhancementpp,strangeness_pPb1,strangeness_pPb2,strangeness_PbPb,strangeness_inJet1,strangeness_inJet2}.

The particle identification is based on the energy loss per unit of track length (\dEdx) measured by the TPC. Protons and pions are identified by requiring that their measured \dEdx is within 5$\sigma_{\mathrm{d}E/\mathrm{d}x}$ from the expected average calculated using the Bethe–Bloch formula, where $\sigma_{\mathrm{d}E/\mathrm{d}x}$ is the \dEdx resolution.

\subsection{Signal extraction}
\label{subsec:SignalExtraction}

\begin{figure}[ht!]
    \centering
    \subfigure[]{\includegraphics[width = 0.48\textwidth]{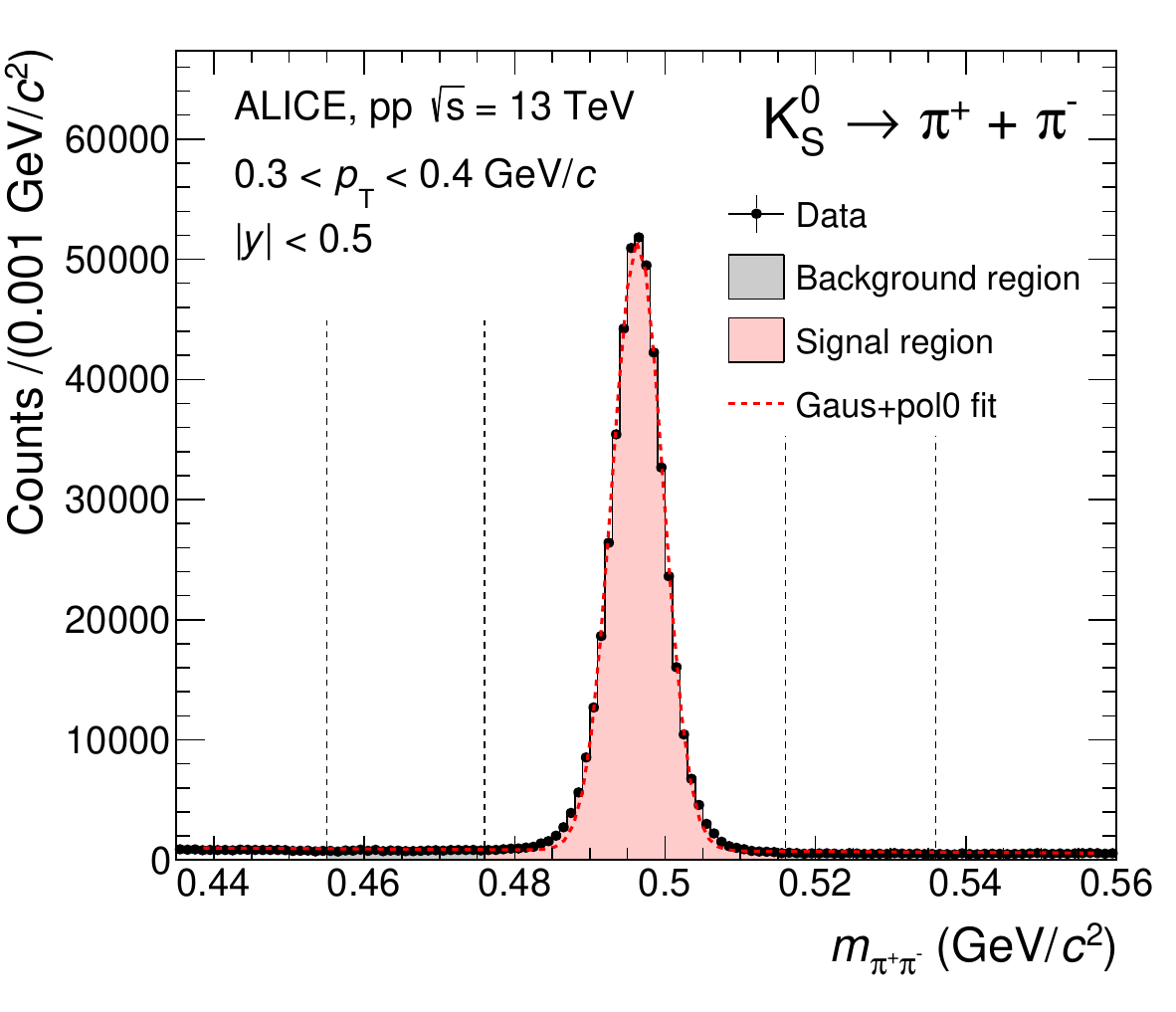}}
    \subfigure[]{\includegraphics[width = 0.48\textwidth]{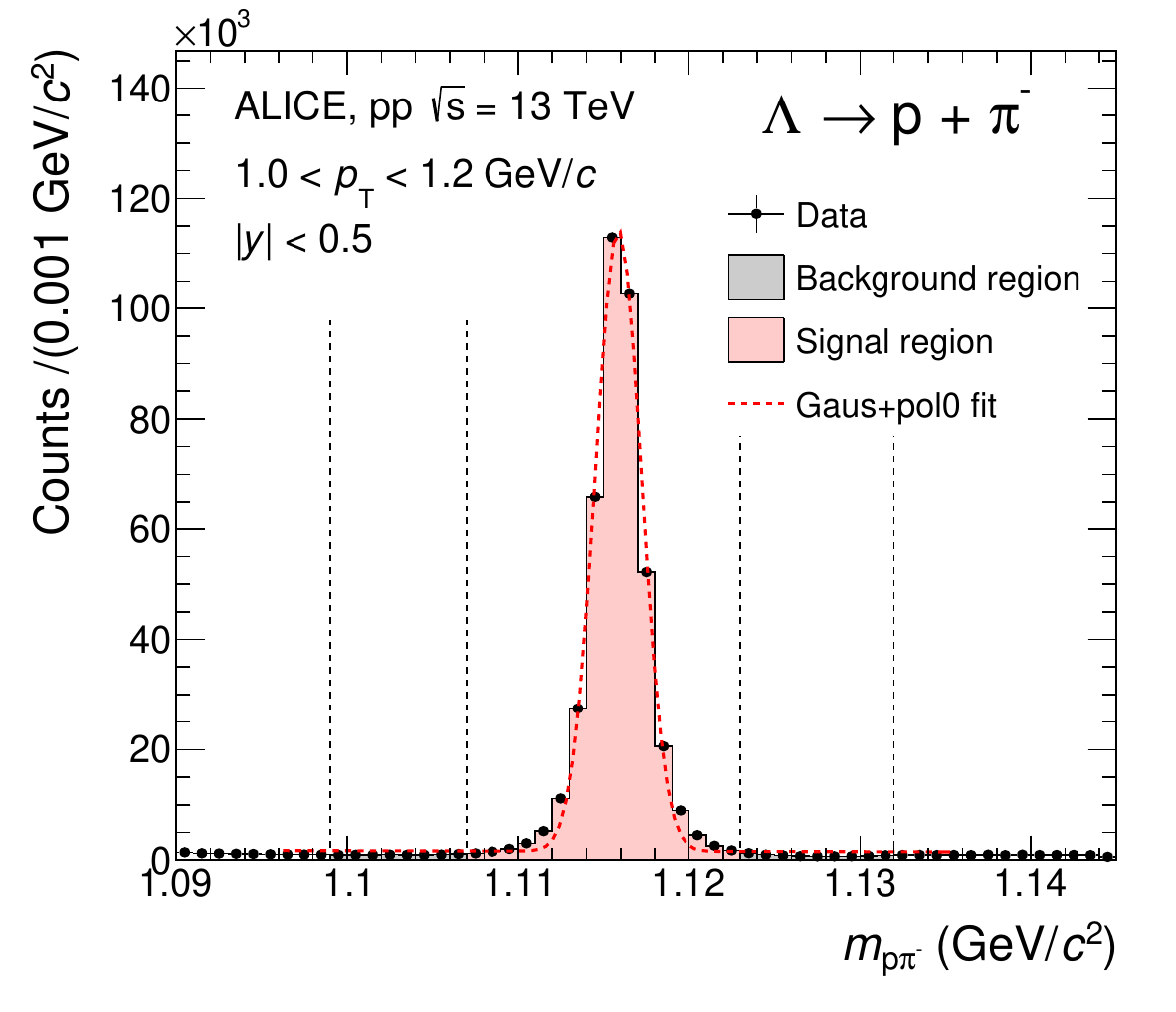}}
   \subfigure[]{\includegraphics[width = 0.48\textwidth]{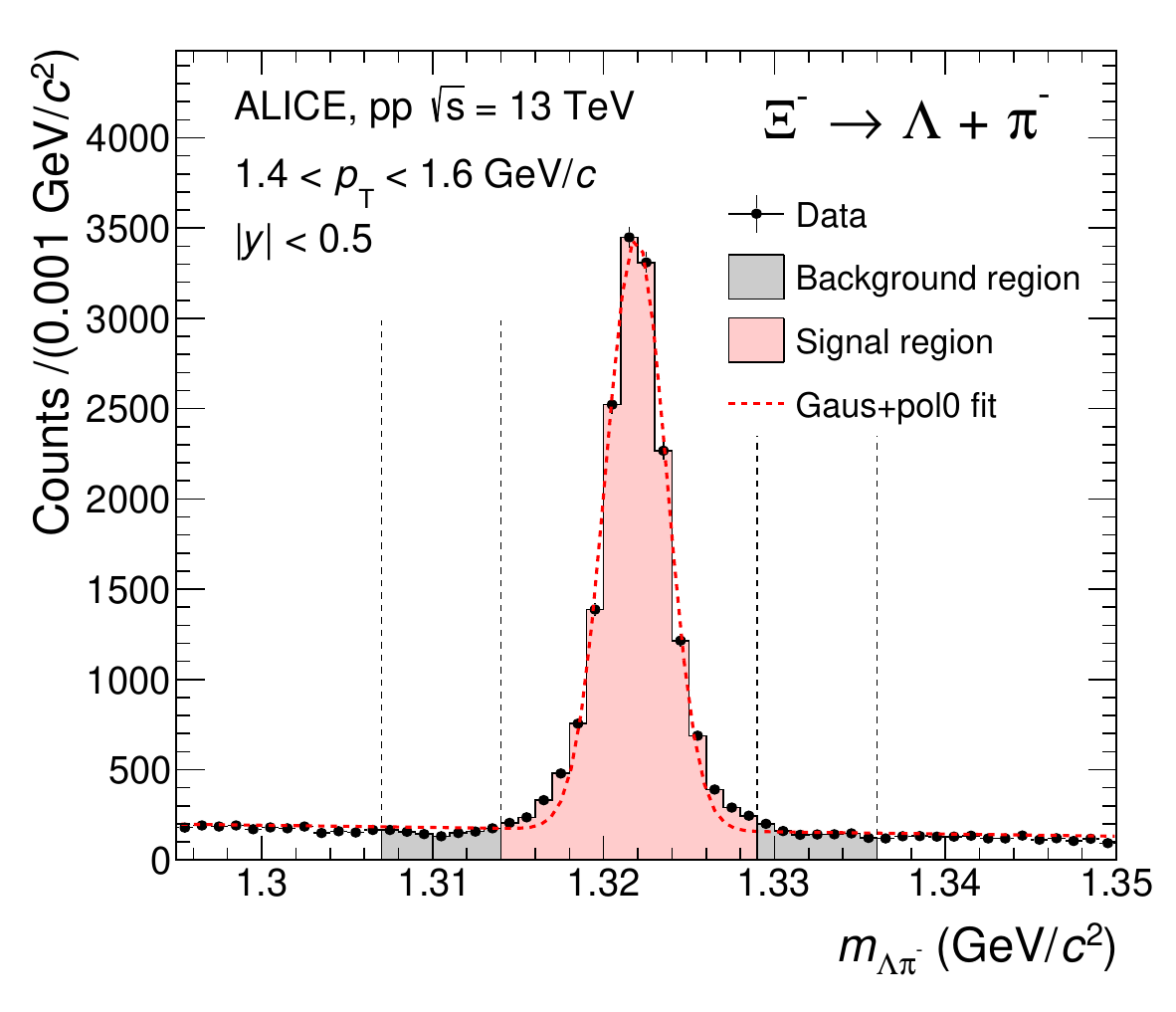}}
    \caption{Invariant mass distributions for \kzero (a), $\Lambda$ (b), and $\Xi^{-}$ (c) in different \pt intervals for the inclusive $\rm{INEL}>0$ class. The candidates are reconstructed in the rapidity interval $|\textit{y}| < 0.5$. The red and grey areas delimited by the short-dashed lines are used for signal extraction in the bin counting procedure. The red dashed lines represent the fit to the invariant mass distributions used to define signal and background regions.}
    \label{fig:invmass}
\end{figure}

The $\rm K_{S}^0$, $\Lambda$, and $\Xi^{\pm}$ raw yields are extracted in different intervals of strange-hadron \pt from fits to the invariant mass distributions of their decay products. The mass distributions were first fitted with a Gaussian function, for modelling the signal, and a linear function to model the background. The peak region is defined within $\pm$ $ 6\mathit{\sigma}$ for $\rm V^{0}$s and $\pm$ $ 4\mathit{\sigma}$ for cascades with respect to the Gaussian mean extracted in each $p_{\textrm{T}}$ interval, being $\mathit{\sigma}$ the standard deviation of the Gaussian function. Adjacent background bands, covering a mass interval as wide as the peak region, are defined on both sides. The symmetric background bands are well reproduced through a linear function, allowing signal extraction through a bin counting procedure that subtracts background counts from the signal region.
The purity of the strange hadron candidate samples, defined as the ratio between the signal and the total number of candidates in the peak region, is larger than 0.9 for \kzero, \lmb and \almb, and larger than 0.8 for \X and \Ix.
Examples of invariant mass distributions and fit functions used for the signal extraction are shown in Fig.~\ref{fig:invmass} for $\rm K_{S}^0$, $\Lambda$, and $\Xi^{-}$ in different \pt intervals.

\subsection{Efficiency and secondary \texorpdfstring{$\Lambda$}{} corrections}
\label{subsec:EfficiencyFeeddown}

The raw \pt spectra of $\rm K_{S}^0$, $\Lambda$(\almb), and \X(\Ix) are corrected for the reconstruction efficiency and the feed-down contribution from higher-mass states. Only the $\Lambda$(\almb) are found to be affected by a significant contamination from feed-down due to the decay of charged \X(\Ix) and neutral $\Xi^{0}$. The contamination from secondary $\rm K_{S}^0$, $\Lambda$(\almb), and \X(\Ix) originating from interactions of primary particles and the detector materials is found to be negligible. All corrections are calculated using Monte Carlo (MC) simulations based on the PYTHIA 8 event generator. The interactions of the generated particles with the experimental apparatus are modelled by GEANT4~\cite{GEANT4}. To reduce the statistical uncertainties on the reconstruction efficiency of cascades without the need to simulate too many events, dedicated MC productions enriched by an injected sample of cascades were used, where one charged $\Xi$ and one charged $\Omega$ are added into each PYTHIA 8 event.
The reconstruction efficiencies of $\rm K_{S}^0$, $\Lambda$(\almb), and \X(\Ix) vary from approximately 2$\%$ at low \pt to about 30$\%$ at high \pt. They are found to be independent of the selection class within 2$\%$. For this reason, the MB efficiency is applied for all classes and a 2$\%$ systematic uncertainty is assigned. In order to estimate the contamination from secondary $\Lambda$(\almb) the measured \X(\Ix) spectra are used. The fraction of secondary $\Lambda$(\almb) particles in the measured spectrum varies between 10$\%$ and 20$\%$, depending on \pt and multiplicity. The input \pt distributions of the injected particles are corrected using multiplicity and \pt-dependent weights calculated as the ratios of measured and injected \pt spectra in the simulation. A Lévy--Tsallis parametrisation~\cite{Tsallis:1987eu} is used to describe the measured \pt shape. The parameters of the Lévy--Tsallis functions are obtained in the first iteration using the efficiency calculated based on the default \pt distributions. The obtained parametrisations in different multiplicity intervals are then used to re-weight the input shape in MC and these updated efficiencies are used to correct the reconstructed spectra. This iterative procedure converges already after two iterations. Finally, in order to achieve a proper normalisation of the spectra, corrections for event loss and signal loss due to the applied event selection criteria are applied to the measured spectra.

The \pt-integrated yields of strange hadrons are then extracted from the transverse momentum spectra, using extrapolations for the unmeasured regions. For this purpouse, the \pt-distributions are fitted using a Lévy--Tsallis parametrisation, which best describes the individual spectra for all particles over the full \pt range.

%% file: Chapters/systematics.tex
\section{Systematic uncertainties}
\label{sec:SystematicUncertainties}
Several sources of systematic uncertainties affecting the measured strange-hadron $p_{\rm T}$-differential yields are considered. 
The different contributions for three representative \pt values are summarised in Table~\ref{tab:SystSpectraSummary} for the INEL$>$0 sample.

To evaluate the systematic uncertainty associated with a given selection, the analysis is repeated by varying the selection criteria on that specific variable within defined ranges. The results are then compared to the ones obtained with the standard set of cuts.
The systematic uncertainty due to possible imperfections in the Monte Carlo simulations used to determine the acceptance and efficiency correction factors was estimated by repeating the analysis varying all track and topological selections. The selection criteria were varied within ranges that led to a maximum variation of $\pm10\%$ in the raw signal yield, similar to the approach used in Refs.~\cite{strangenessEnhancementNature,strangenessEnhancementpp}. The corrected yields were calculated for each variation and compared with those obtained with the default selections. Only variations giving results differing from the default ones by more the $1\sigma$ of the statistical uncertainty were considered to define the systematic uncertainty, following the prescription in Ref.~\cite{barlow2002systematic}. The uncertainty was found to range from about $1\%$ to about $4\%$ depending on the hadron species and \pt.

The uncertainty related to the PID selections was evaluated by varying the $\textrm{d}E/\textrm{d}x$ requirement between 4$\sigma$ and 7$\sigma$.
This selection is particularly important to reduce the combinatorial background in the strange baryon invariant mass distribution. 
The uncertainty was found to be at most 1$\%$ for all particle species.

\def\arraystretch{1.2}
  \begin{table}[htbp!]
    \caption{Main sources of systematic uncertainties (expressed in $\%$) of the \pt differential yields, reported for low, intermediate, and high $p_{\rm T}$.
    These values are calculated for the INEL$>$0 data sample. Results in other classes are further affected by an uncertainty originating from the class selection dependence of the efficiency (2$\%$) and, in the case of the $\Lambda$ and \almb, of the feed-down contributions (2$\%$).}
    \small
    \centering
    \begin{tabular}{ c | c  c  c | c  c  c | c  c  c  }
     \hline
     \hline
      Hadron & \multicolumn{3}{ c }{\textbf{\kzero}} &  \multicolumn{3}{| c |}{\textbf{$\Lambda+\overline{\Lambda}$}} &  \multicolumn{3}{ c }{\textbf{$\Xi^{-}+\overline{\Xi}^{+}$}} \\
      \pt (\rm{GeV/\textit{c}}) & $\approx0.5$ & $\approx4.8$ & $\approx9.0$ & $\approx1.0$ & $\approx3.5$ & $\approx7.0$ & $\approx1.3$ & $\approx2.8$ & $\approx4.7$ \\
      \hline
      \hline
      Topological and track & 1.7 & 2.4 & 2.3 & 3.0 & 3.0 & 4.3 & 3.3 & 1.1 & 2.0 \\
      TPC $\textrm{d}E/\textrm{d}x$ & 0.1 & 0.1 & negl. & 0.3 & 0.1 & 0.6 & 0.3 & 0.1 & negl. \\
      Competing $\mathrm{V}^{0}$ & 0.1 & 0.4 & negl. & 1.1 & 0.3 & negl. & n.a. & n.a. & n.a. \\
      Proper lifetime & 0.1 & negl. & negl. & 2.9 & 2.6 & negl. & 0.7 & 0.2 & 0.6 \\
      Signal extraction & 0.5 & 0.5 & 2.2 & 0.3 & 0.7 & 1.2 & negl. & 0.6 & 0.6 \\
      Feed-down & n.a. & n.a. & n.a. & 1.0 & 1.0 & 1.0 & n.a. & n.a. & n.a. \\
      $\overline{\rm p}$ abs. cross sect. & n.a. & n.a. & n.a.& 0.2 & 0.2 & 0.2  & 0.3 & 0.4 & 0.4 \\  
      In-bunch (IB) pileup & 1.6 & 2.5 & 2.5 & 2.0 & 2.9 & 2.9 & 2.0 & 2.0 & 2.9 \\
      OOB pileup & 0.2 & 0.8 & 2.6 & 0.6 & 2.1 & 2.9 & 0.4 & 1.0 & 2.5 \\
      Material budget & 1.1 & 0.5 & 0.5 & 1.4 & 0.8 & 0.8 & 2.9 & 1.5 & 0.6 \\
      \hline
      Total & 2.7 & 3.6 & 5.0 & 5.1 & 5.5 & 5.9 & 4.9 & 2.9 & 4.4 \\
      \hline
      \hline
    \end{tabular}
    \label{tab:SystSpectraSummary}
  \end{table}
  
The contribution from the competing V$^{0}$ decay rejection was studied by removing entirely this condition for $\Lambda$ and $\overline{\Lambda}$ and by varying the mass window down to $3~{\rm MeV}/c^{2}$ and up to $5.5~{\rm MeV}/c^{2}$ for \kzero.
It resulted in a deviation on the corrected yields of at most 3$\%$ for $\Lambda+\overline{\Lambda}$ and $1\%$ for \kzero.

The systematic contribution related to the selection on the proper lifetime of the candidate was computed by varying the requirements between 2.5 and 5 $c\tau$ for strange baryons and between 5 and 15 $c\tau$ for \kzero.
The statistically significant deviations were found to be at most $3\%$ for $\Lambda+\overline{\Lambda}$ and negligible ($<1\%$) for \kzero and $\Xi^{-}+\overline{\Xi}^{+}$.

The stability of the signal extraction method was checked by varying the widths used to define the signal and background regions in the invariant mass distributions in terms of the number of sigmas of the signal peak.
Moreover, the raw counts are extracted using a fitting procedure for the background contribution and compared to the standard ones computed using a bin counting technique.
An uncertainty ranging up to 2$\%$ depending on \pt is found for V$^{0}$s and cascades. 

A 2$\%$ uncertainty is added to account for possible variations of the reconstruction efficiency with the class selections (Sec.~\ref{subsec:EfficiencyFeeddown}).

The \lmb and $\overline{\Lambda}$ \pt spectra are also affected by an uncertainty coming from the feed-down correction, which accounts for the description of the $\Xi^{\pm}/\Xiz$ ratio in the MC.
The latter was considered by calculating the feed-down fraction assuming $\Xi^{\pm}/\Xiz~=~1$ or using the ratio provided by the Monte Carlo.
The feed-down contribution to the systematic uncertainties was at most 1$\%$.
An additional 2$\%$ uncertainty, related to the systematic uncertainty on the efficiency discussed above, is added to account for possible variations of the feed-down contribution with the class selections.

The systematic uncertainty on the \lmb (\almb) and $\Xi^{\pm}$ yield due to the (anti-)proton absorption in the detector material is estimated by varying the default inelastic cross section of (anti-)protons implemented in GEANT4 by the corresponding experimental uncertainties~\cite{antiprotonAbsorption}. This contribution is found to be less than 1$\%$ for strange baryons. 

Pile-up collisions occurring within the same bunch crossing are removed by rejecting events with multiple vertices reconstructed with SPD tracklets.
The effect of residual contamination from in-bunch pile-up events was estimated in Ref.~\cite{strangenessEnhancementpp} by varying the pile-up rejection criteria. In this analysis, the same systematic uncertainties are used.

The contribution from the out-of-bunch pile-up rejection was evaluated by changing the matching scheme of V$^{0}$ and cascade daughters using the ITS and TOF detectors.
For this purpose, the following configurations were considered: matching of at least one decay track of the reconstructed (multi-)strange hadron with the ITS (TOF) detector below (above) $2~{\rm GeV/}c$, ITS matching of at least one decay track in the full \pt range.
Half of the maximum variation between these configurations and the standard selection was taken as a systematic contribution, which was found to increase with transverse momentum up to $3\%$ for all particle species.

The systematic uncertainty due to the limited knowledge of the ALICE material budget is estimated using two MC productions: one with a value of the material density obtained from the geometrical model of the ALICE experiment implemented in the simulation, and another where the material density was modified locally to match the measurement obtained using photon conversions~\cite{Acharya_2023}. The material budget uncertainty is calculated as the relative difference between the efficiencies obtained with these two MC productions.

Most of the sources of systematic uncertainties considered are fully correlated across the classes of events defined using the V0M and SPDClusters estimators since they determine a yield variation that does not depend on the specific event class.
In this analysis, to illustrate the evolution of the production of strange hadrons in the multi-differential classes and reduce the systematics on the final results, the yield ratios to the average value, measured in the inclusive INEL$>$0 pp sample, are considered:
\begin{displaymath}
\label{eq:yieldratio}
   \langle h \rangle / \langle h \rangle_{\rm{INEL}>0}
\end{displaymath}
where $h$ is the \pt integrated yield ($\textrm{d}N/\textrm{d}y$) of strange hadrons.
In order to determine the fraction of uncorrelated uncertainty, the full analysis chain, up to the extraction of integrated yields, is repeated for each event class by varying the $\rm V^{0}$ and cascade selection criteria and comparing the results with the ones obtained with the default set of cuts. The relative deviation of yields is then compared to the corresponding one obtained using the $\rm INEL>0$ sample.
A fully correlated uncertainty is characterised by a ratio of relative deviations between the event class and the inclusive $\rm INEL>0$ sample that is consistent with unity. 
Deviations from unity are considered uncorrelated components of the uncertainty.
The fraction of uncorrelated systematic uncertainties due to the analysis selections ranges between 0.1$\%$ and 1$\%$ for \kzero, between 0.1$\%$ and 1.5$\%$ for $\Lambda+\overline{\Lambda}$, and between 1$\%$ and 3$\%$ for $\Xi^{-}+\overline{\Xi}^{+}$.
A similar approach is applied to estimate the fraction of the systematic uncertainty uncorrelated across selected event classes due to the choice of the fit function for the extrapolation procedure.
The yields obtained using the extrapolation from an alternative function are compared to the default one, Lévy-Tsallis, in a given selection class. Then, these are compared to the same results obtained in the inclusive $\rm INEL>0$ class. The fit of the spectra is repeated using five alternative functions (Blast-Wave, Boltzmann, Bose--Einstein, $m_{\text{T}}$-exponential, Fermi--Dirac). 
Since these alternative functions do not describe the full $p_{\textrm{T}}$-distribution, in this case, the fit range was reduced to obtain a good description of the fitted part of the spectrum.
The fraction of uncorrelated systematic uncertainties due to the choice of the fit function ranges between 0.5$\%$ and 4$\%$ for $\Lambda+\overline{\Lambda}$, and between 1$\%$ and 9$\%$ for $\Xi^{-}+\overline{\Xi}^{+}$. No uncertainty on the extrapolation is considered for the \kzero meson since the spectra are measured down to \pt = 0.

The systematic uncertainty associated with the average $ZN$ value was studied by comparing the energy measured in the three data-taking periods (2015, 2017, and 2018). The measurement was repeated in different double-differential classes defined using SPDClusters and V0M estimators. The error associated with the $\langle ZN \rangle$ value was defined as the largest difference of each period with respect to the mean value, which is found to be at the level of $\sim3\%$.

%% file: Chapters/results.tex
\section{Results and discussion}
\label{sec:results}
\begin{figure}[!ht]
	\centering
	\includegraphics[width=0.94\textwidth]{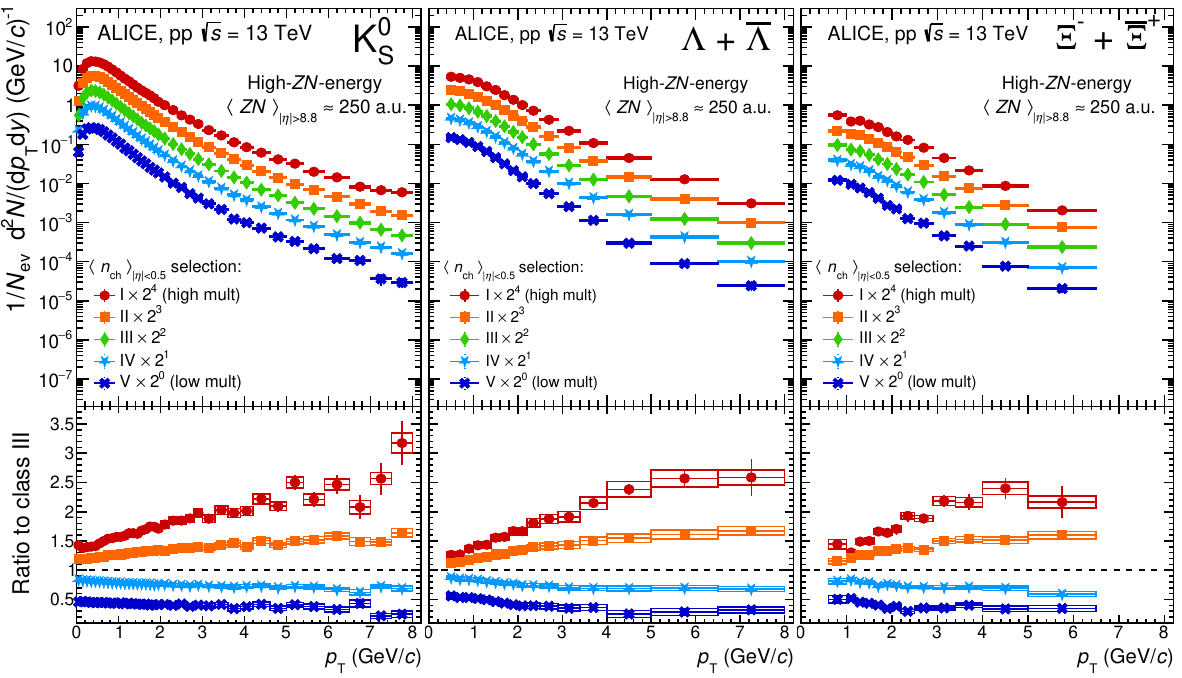}
	\caption{Transverse momentum distribution of \kzero (left), $\Lambda$ (middle), and $\Xi$ (right) in the high-$ZN$-energy selections (SPDClusters+V0M classes). The bottom panels show the ratios of the spectra in the different classes to the central one. The spectra are scaled by different factors to improve the visibility.}
	\label{fig:spectrazn}
\end{figure}

\begin{figure}[!ht]
	\centering
	\includegraphics[width=0.94\textwidth]{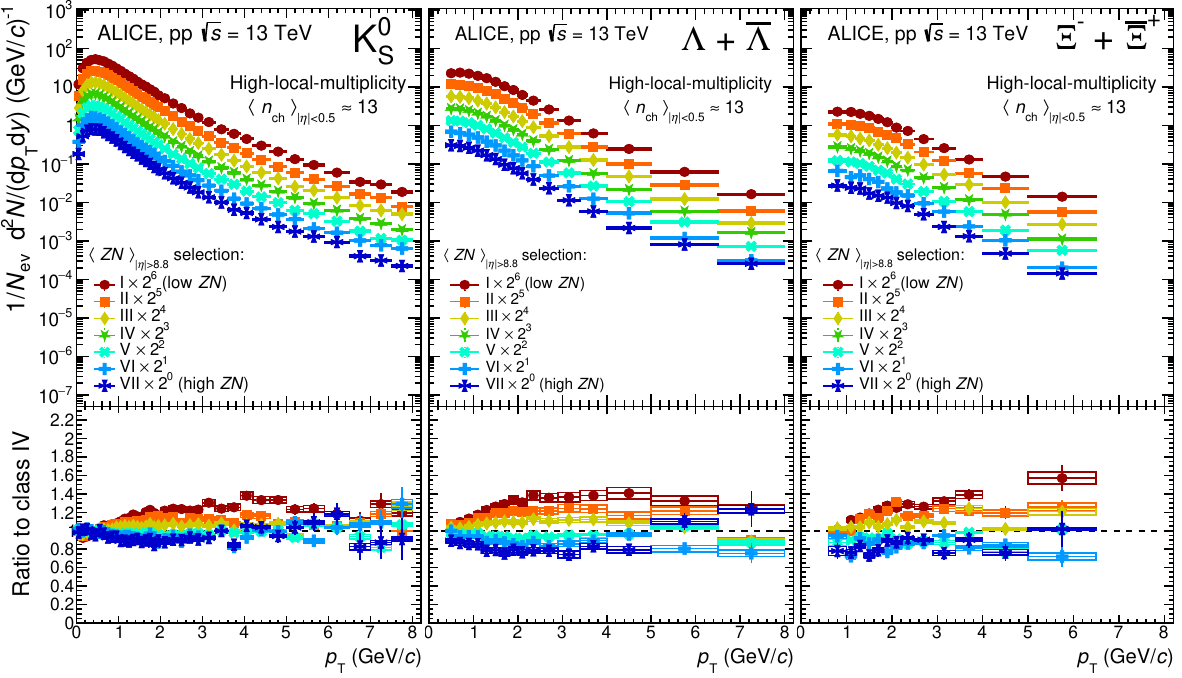}
	\caption{Transverse momentum distribution of \kzero (left), $\Lambda$ (middle), and $\Xi$ (right) in the high-local-multiplicity selections (SPDClusters+V0M classes). The bottom panels show the ratios of the spectra in the different classes to the central one. The spectra are scaled by different factors to improve the visibility.}
	\label{fig:spectranch}
\end{figure}

In the following, the sum of particles and anti-particles, $\Lambda+\almb$, $\X+\Ix$, will be referred to as \lmb and $\Xi$, respectively.
The fully-corrected $p_{\rm T}$-differential yields for all particle species are displayed in Fig.~\ref{fig:spectrazn} for the five multiplicity classes defined with the high-$ZN$-energy classification and in Fig.~\ref{fig:spectranch} for the seven event classes with the high-local-multiplicity classification.
Qualitatively similar results are obtained for the complementary selections: low-$ZN$-energy and low-local-multiplicity.
In events with defined $ZN$ energy, the \kzero, $\Lambda$, and $\Xi$ spectra become harder and the yields increase as the local multiplicity increases. This is clearly visible when looking at the ratio of the distributions with respect to the central class spectrum, shown in the bottom panel of Fig.~\ref{fig:spectrazn}.
On the other hand, as displayed in Fig.~\ref{fig:spectranch}, for similar values of midrapidity multiplicity, the \pt spectra of strange hadrons are found to be similar in the different classes of events selected on the basis of different forward energy deposits.
The dynamics in terms of the transverse momentum observed with our two-dimensional approach suggests that the average \pt of (multi-)strange hadrons is strongly correlated with the local multiplicity production.
This indicates that, once the activity at forward rapidity is fixed, the increase of strangeness production with multiplicity at midrapidity is driven by harder processes, in line with Fig.~\ref{fig:meanpt}.
To illustrate the evolution of strange hadron production in the different event classes, the \pt-integrated yield ratios to the charged-particle multiplicity divided by the value measured in the inclusive $\rm INEL > 0$ pp sample are considered:
\begin{equation}
\label{eq:yieldratios}
    \frac{\langle h \rangle / \langle h \rangle^{\rm{INEL}>0}}{\langle n_{\rm{ch}} \rangle_{|\eta|<0.5}/ \langle n_{\rm{ch}} \rangle_{|\eta|<0.5}^{\rm{INEL}>0}} \quad ,
\end{equation}
where $h$ represents the particle yield per rapidity unit ($\textrm{d}N/\textrm{d}y$). The uncertainties on the ratios are propagated assuming the two variables to be uncorrelated.
In the following, the quantity in Eq.~\ref{eq:yieldratios} will be referred to as ``self-normalised yield ratios''.
The yield per charged particle ($h/n_{\rm ch}$) is a good proxy for the ratio of strange-to-charged-pion yields ($h/\pi$).

\subsection{Standalone analysis results}

\begin{figure}[!hbt]
    \centering
    \includegraphics[width = 0.95\textwidth]{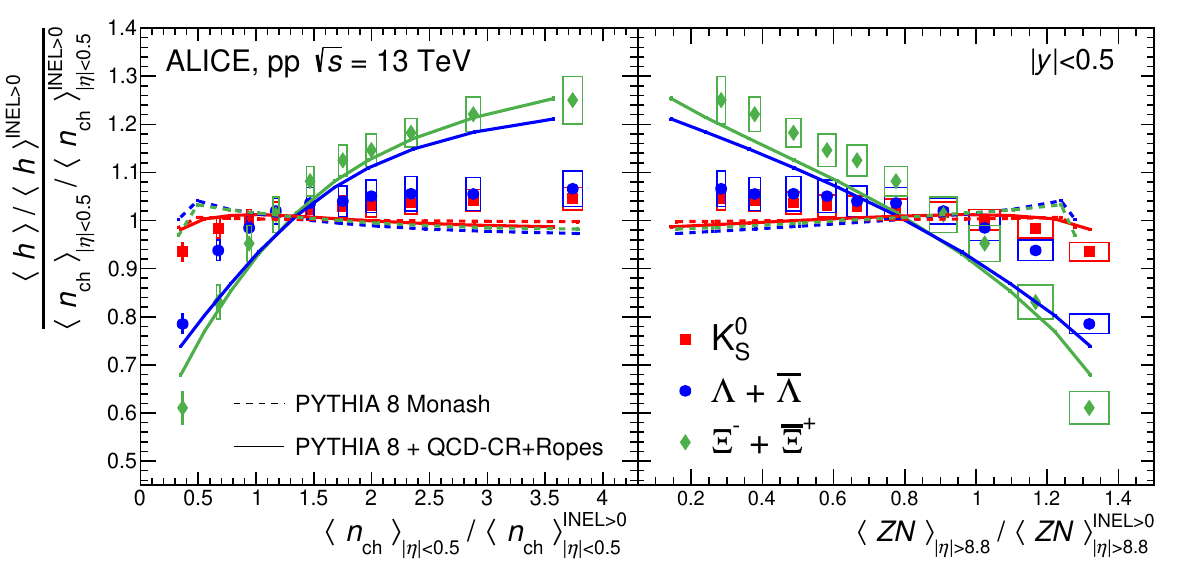}
    \caption{Self-normalised yield ratios of \kzero, \lmb, and $\Xi$ in pp collisions at $\sqrt{s} = 13~\rm TeV$ in the standalone selection (V0M classes). The ratios are shown as a function of the average charged-particle multiplicity $\langle n_{\rm ch} \rangle$ (left) and the forward energy $\langle ZN \rangle$ (right) self-normalised to the minimum bias ($\rm INEL>0$) value. Statistical and total systematic uncertainties are shown by error bars and boxes, respectively.}
    \label{fig:selfyieldsstandalone}
\end{figure}

The self-normalised yield ratios of \kzero, \lmb, and $\Xi$ measured in the event classes defined with the standalone selections are shown in Fig.~\ref{fig:selfyieldsstandalone} as a function of the average charged-particle multiplicity $\langle n_{\rm ch} \rangle$ and of the average energy measured in the neutron calorimeter $\langle ZN \rangle$, self-normalised to their $\rm INEL > 0$ value.
The left panel shows that the strange hadron yields per charged particle increase as a function of the charged-particle multiplicity at midrapidity.
The enhancement is larger for $\Xi$ multi-strange baryons than for \lmb and \kzero strange hadrons, showing a hierarchy with the particle strangeness content.
The \lmb baryon and \kzero meson ratios are compatible within uncertainties, except for the lowest multiplicity interval.
These observations are consistent with what was observed in previous ALICE publications~\cite{strangenessEnhancementNature,strangenessEnhancementpp}.
To explore the correlation of strangeness production at midrapidity with the very-forward energy, the ratios are also displayed as a function of the self-normalised $ZN$ signal in the right panel of Fig.~\ref{fig:selfyieldsstandalone}.
Strange hadron production per charged particle is found to increase with decreasing forward energy detected in the $ZN$. This observation demonstrates a positive correlation of strangeness production with the effective energy.
The results are compared with MC simulations based on the PYTHIA 8 event generator.
The Monash tune does not reproduce the strangeness enhancement observed in the data, showing a flat trend as a function of multiplicity and leading energy for all particle species, as already discussed in previous publications~\cite{strangenessEnhancementpp, Acharya_2020}.
The QCD-CR+Ropes tune strongly improves the agreement of the model with the data points for the $\Xi$ baryon.
However, it overestimates the production of $\Lambda$ baryons per charged particle at high multiplicity.
In this model, the enhancement observed for the $\Xi$ and $\Lambda$ baryons is found to be similar while no enhancement is foreseen for the \kzero meson, missing the increasing trend observed in the data.
The hadronisation mechanisms implemented in the QCD-CR+Ropes tune result in an enhanced production of strange baryons with respect to strange mesons, failing to reproduce the hierarchy with the strangeness content observed in the data.
This observation is further confirmed when looking at the predictions of the QCD-CR+Ropes tune for protons, which show a rising trend with increasing multiplicity not observed in the data~\cite{Acharya_2020}.
It is important to note that, even if this tune reproduces the $\Xi/n_{\rm ch}$ enhancement in the standalone event class, it is known that PYTHIA does not perfectly reproduce the transverse momentum spectral shapes of strange hadrons~\cite{loizides2021apparent}.

\subsection{Strangeness production in events with defined leading energy and multiplicity}
\begin{figure}[htp]
    \centering
    \includegraphics[width = \textwidth]{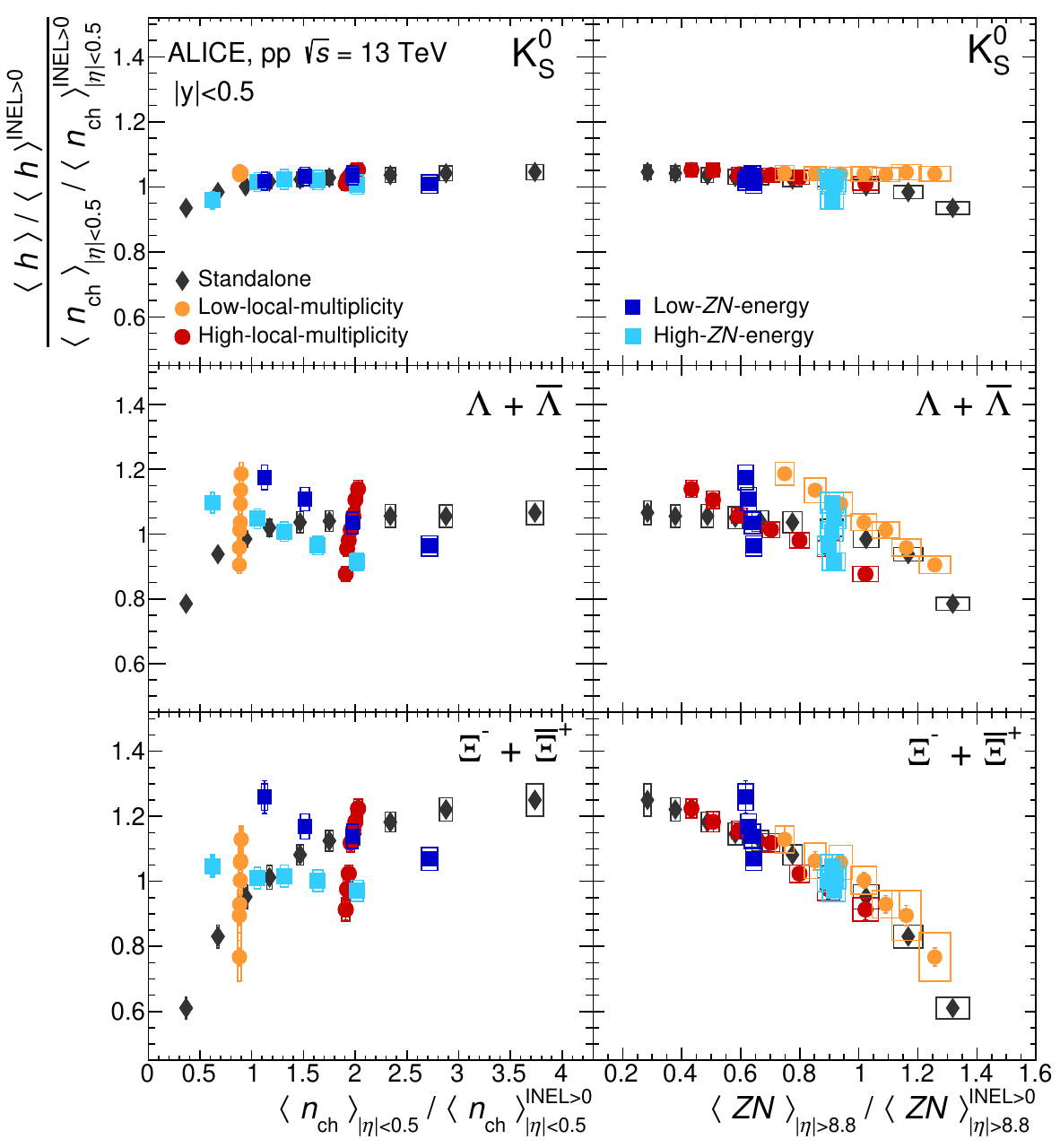}
    \caption{Self-normalised yield ratios of \kzero, \lmb, and $\Xi$ in pp collisions at $\sqrt{s} = 13~\rm TeV$ in the high/low multiplicity and high/low $ZN$ selections (V0M+SPDClusters classes). The ratios are shown as a function of the average charged-particle multiplicity $\langle n_{\rm ch} \rangle$ (left) and the forward energy $\langle ZN \rangle$ (right), self-normalised to the minimum bias ($\rm INEL>0$) value. Statistical and total systematic uncertainties are shown by error bars and boxes, respectively. }
    \label{fig:selfyieldsmultdiff}
\end{figure}

The dependence of strange hadron production on the charged-particle multiplicity can be further investigated in events with similar average forward energy measured in the $ZN$.
For this purpose, the multiplicity classes defined for the high-/low-$ZN$-energy categories are considered.
Similarly, the dependence on the leading energy can be studied in events with similar average charged-particle multiplicity produced at midrapidity using the high-/low-local-multiplicity event classes.
The self-normalised yield ratios obtained in these selections for all particle species are displayed in Fig.~\ref{fig:selfyieldsmultdiff} as a function of the charged-particle multiplicity (left) and $ZN$ energy (right), self-normalised to their $\rm INEL > 0$ value.
The standalone selection is also shown for comparison.
Once events with defined leading energy are considered (azure and blue squares), the strange baryon enhancement with multiplicity is no longer observed. Instead, the $\Lambda$ and $\Xi$ yield ratios show a mild anti-correlation with particle production at midrapidity.
For the \kzero meson a mild-to-no increase with the midrapidity multiplicity is observed for events with similar $ZN$ energy. In particular, the trend for \kzero is compatible with the results of the standalone analysis.
The decreasing trend with multiplicity observed for strange baryons could be explained by introducing a simple energy conservation argument: at fixed effective energy, as the charged-particle multiplicity increases, the remaining energy available for the production of heavier strange hadrons decreases.
In this case, the production yield of strange baryons at midrapidity may be more strongly correlated to the effective energy than the production of light mesons.
Alternatively, the observed anti-correlation could be interpreted considering that events with similar effective energy may be characterised by different topologies in terms of hard scattering processes, associated with the production of jets, in line with the hardening of the spectra observed in Fig.~\ref{fig:spectrazn}.
In particular, the presence of one or more jets at midrapidity may result in a large local production of charged particles.

For similar midrapidity multiplicities (orange and red circles), the self-normalised yield ratios of $\Lambda$ and $\Xi$ increase with decreasing energy deposited in the neutron calorimeters.
For $\Xi$ baryons, the scaling with the $ZN$ energy is observed to be compatible, within uncertainties, with the one observed for the standalone classification for both the low- and high-local-multiplicity selections.
In contrast, for $\Lambda$ baryons, the dependence on the forward energy is not common among the different types of classifications.
The $\Lambda$ ratios in the high-local-multiplicity classes are found to be systematically lower than those in the low-local-multiplicity classes at similar values of $ZN$ energy. This observation implies that, in events with similar leading energy, a smaller production of $\Lambda$ baryons per charged particle is observed in events with larger values of multiplicity at midrapidity. This is in agreement with the decrease of $\Lambda$ yield with increasing midrapidity multiplicity observed in the fixed leading energy classes, discussed above.
It is worth noting that in terms of charged-particle multiplicity the high-$ZN$-energy classification (azure squares) spans a $n_{\rm ch}$ range that encompasses the values of the low- and high-local-multiplicity classes.
In this classification, while the $\Xi$ production is almost constant within the classes, showing no dynamic once $ZN$ is fixed, the $\Lambda$ self-normalised yields vary by about 20$\%$ from the lowest to the highest $n_{\rm ch}$ values.
Notably, the $\Lambda$ production per charged particle in the high-/low-local-multiplicity classes reaches overall higher values with respect to the highest values obtained with the standalone selections.
The \kzero meson, on the other hand, shows very mild-to-no dependence on the $ZN$ energy once the midrapidity activity is fixed.
Comparing the results for the different particle species provides interesting inputs on the correlation of strange hadron production with the leading energy once the charged particle multiplicity is defined.
The larger effect observed for $\Lambda$ with respect to \kzero indicates a stronger correlation of strange baryon production with the leading energy with respect to mesons, given the two hadrons have the same strangeness content.
This observation is further supported by comparing the results of $\Lambda$ to $\Xi$ yields, which show a compatible relative increase at fixed multiplicity.
However, it is worth noting that the dependence of $\Lambda$ to $\Xi$ yield ratios to the $ZN$ energy differs in these event classes, suggesting that the hadron strangeness content plays a role in the observed behaviour.

\subsection{Comparison to PYTHIA 8}
\begin{figure}[!hbt]
    \centering
    \includegraphics[width = 0.9\textwidth]{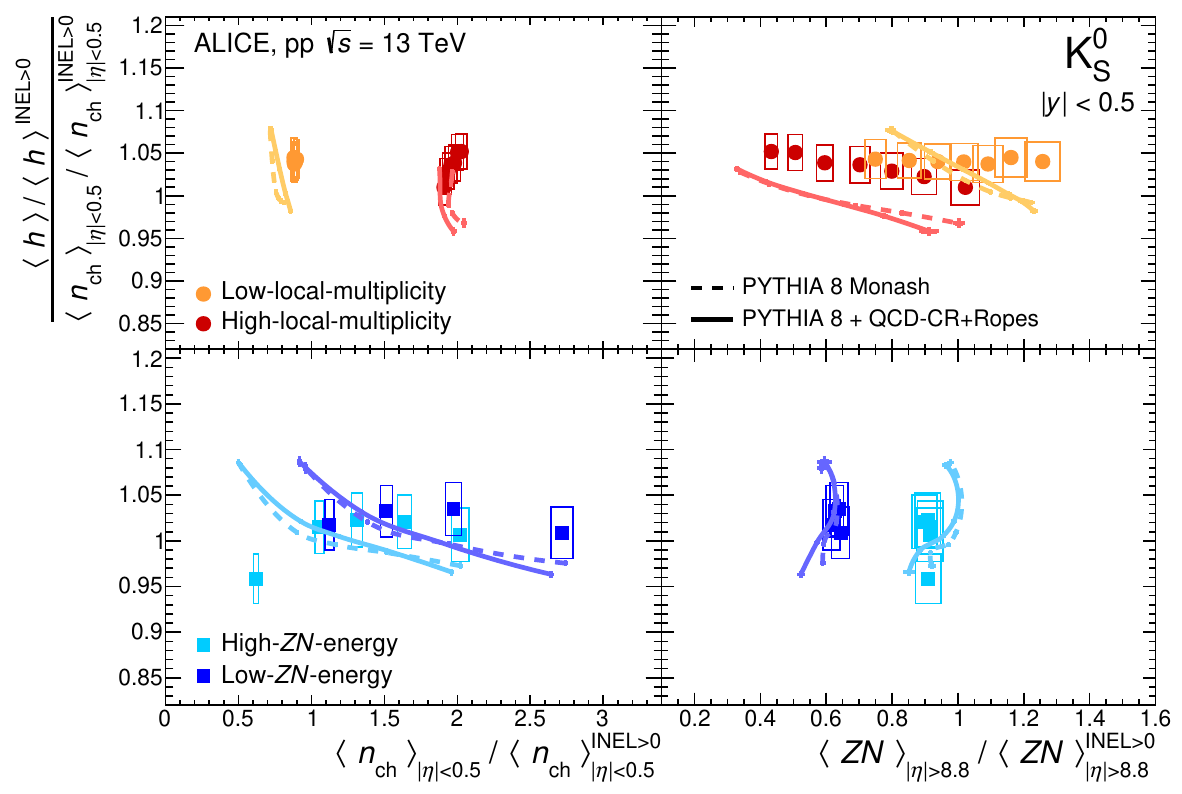}
    \caption{Self-normalised yield ratios of \kzero in pp collisions at $\sqrt{s} = 13~\rm TeV$ in the high-/low-local-multiplicity, high-/low-$ZN$-energy selections compared to PYTHIA 8 Monash and QCD-CR+Ropes predictions. The classes at fixed multiplicity are displayed in the top panels, and the classes at fixed leading energy are displayed in the bottom panels. The results from PYTHIA 8 with Monash and QCD-CR+Ropes tunes are shown with dashed and full lines, respectively. Statistical and total systematic uncertainties are shown by error bars and boxes, respectively.}
    \label{fig:modelsK0s}
\end{figure}

\begin{figure}[!hbt]
    \centering
    \includegraphics[width = 0.9\textwidth]{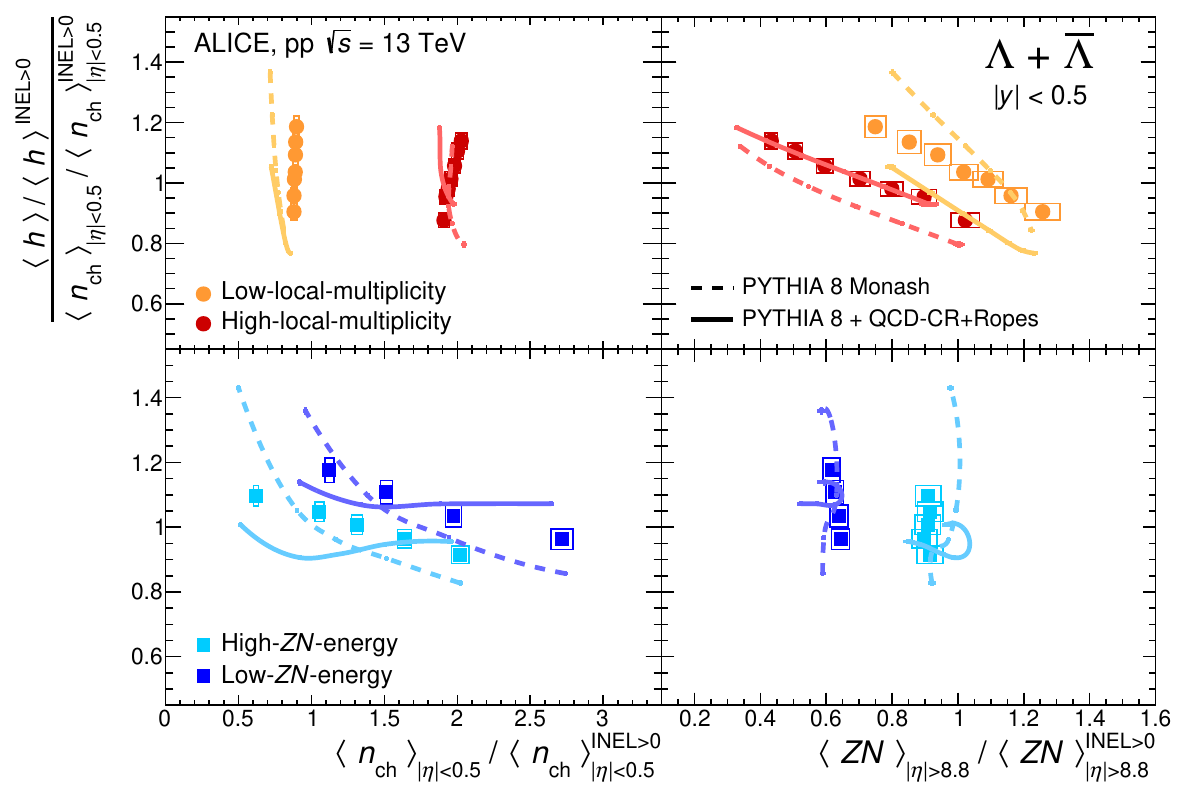}
    \caption{Self-normalised yield ratios of \lmb in pp collisions at $\sqrt{s} = 13~\rm TeV$ in the high-/low-local-multiplicity, high-/low-$ZN$-energy selections compared to PYTHIA 8 Monash and QCD-CR+Ropes predictions. The classes at fixed multiplicity are displayed in the top panels, and the classes at fixed leading energy are displayed in the bottom panels. The results from PYTHIA 8 with Monash and QCD-CR+Ropes tunes are shown with dashed and full lines, respectively. Statistical and total systematic uncertainties are shown by error bars and boxes, respectively.}
    \label{fig:modelsLam}
\end{figure}

\begin{figure}[!hbt]
    \centering
    \includegraphics[width = 0.9\textwidth]{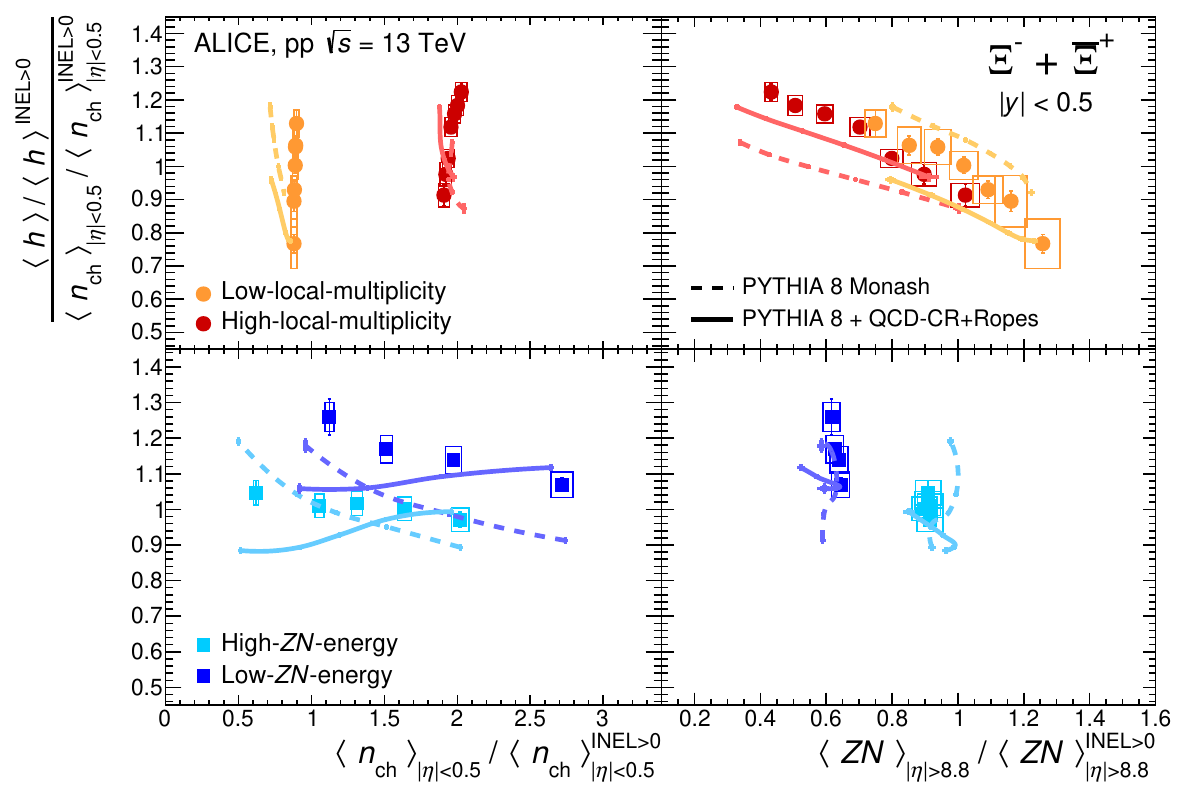}
    \caption{Self-normalised yield ratios of $\Xi$ in pp collisions at $\sqrt{s} = 13~\rm TeV$ in the high-/low-local-multiplicity, high-/low-$ZN$-energy selections compared to PYTHIA 8 Monash and QCD-CR+Ropes predictions. The classes at fixed multiplicity are displayed in the top panels, and the classes at fixed leading energy are displayed in the bottom panels. The results from PYTHIA 8 with Monash and QCD-CR+Ropes tunes are shown with dashed and full lines, respectively. Statistical and total systematic uncertainties are shown by error bars and boxes, respectively.}
    \label{fig:modelsXi}
\end{figure}

The results reported in Fig.~\ref{fig:selfyieldsmultdiff} are compared with MC simulations based on PYTHIA 8 in Figs.~\ref{fig:modelsK0s}, ~\ref{fig:modelsLam}, and~\ref{fig:modelsXi} for \kzero, \lmb, and $\Xi$, respectively.
In particular, Fig.~\ref{fig:modelsK0s} shows the self-normalised yield ratios for the \kzero meson, compared to simulations with PYTHIA 8 with the Monash tune (dashed line) and the QCD-CR+Ropes tune (full line).
The two tunes predict a very similar behaviour for \kzero, suggesting that including improved colour reconnection and ropes in the hadronisation process does not have a big impact on the production of strange mesons.
For events with defined average midrapidity multiplicity (top left and right panels), the two tunes predict a mild and yet significant dependence on the leading energy, in moderate tension with the measured trend.
For similar leading energies (bottom left and right panels), both PYTHIA 8 tunes predict a very mild decrease of the \kzero yield per charged particle with the charged-particle multiplicity, not matching the data at low multiplicity.

Figure~\ref{fig:modelsLam} shows the model comparison to the self-normalised yield ratios for the \lmb baryon in the event classes with high-/low-local-multiplicity and high-/low-$ZN$-energy.
For similar midrapidity multiplicity values, the two tunes predict an increase of \lmb production per charged particle with decreasing $ZN$ energy.
It is worth noting that the PYTHIA 8 Monash tune predicts no strange hadron enhancement in the standalone event selection, however, once the multiplicity is fixed, an increase is observed with decreasing leading energy.
For similar leading energy values, the Monash tune predicts a decrease of the \lmb yield per charged particle with increasing local multiplicity, similarly to what is observed in the data points, but struggling to reproduce the measured values quantitatively.
The QCD-CR+Ropes tune predicts small-to-no dynamics with multiplicity once the leading energy is fixed.

Finally, Fig.~\ref{fig:modelsXi} shows the model comparison to the self-normalised yield ratios for the $\Xi$ baryon for the different event classes.
In this case, the PYTHIA 8 event generator including improved colour reconnection and ropes does an excellent job in reproducing the data points in the standalone selection, as discussed above.
For similar midrapidity multiplicities, the QCD-CR+Ropes tune qualitatively describes the increase of $\Xi$ production per charged particle with decreasing leading energy, also reproducing the universal trend observed in the data points with the $ZN$ energy.
However, the model struggles to reproduce the measured data points quantitatively.
The Monash tune fails to reproduce the strangeness enhancement in the standalone event selection (see Fig.~\ref{fig:selfyieldsstandalone}), however, once the multiplicity is defined, an increase is observed with decreasing leading energy qualitatively similar to the measured trend, although it predicts a difference between the high and low-local-multiplicity classes for the same ZN energy, which is not observed in the data.

Given the anti-correlation between the forward energy and the number of MPIs, the PYTHIA 8 predictions for the \lmb and $\Xi$ baryons may indicate that also at fixed midrapidity multiplicity an increase in strange baryon production is expected at increasing number of MPIs, regardless of the hadronisation mechanism at play. On the other hand, this is not foreseen for strange \kzero mesons.
The largely different predictions for \lmb and \kzero hadrons obtained including the rope formation mechanism in the model suggest that the interplay between MPIs and colour rope hadronisation may have a stronger impact on the enhancement of (strange) baryons, rather than on the enhancement of strangeness itself. This interpretation is further supported by comparing the results of \lmb and $\Xi$ baryons, for which similar trends are predicted by the tune with QCD-CR+Ropes, despite the different strangeness content.

In conclusion, to produce an enhancement of strange baryon yields in PYTHIA 8 the interplay between the MPIs and the hadronisation mechanism appears to be essential, with MPIs increasing the string density and rope formation mechanism effectively enhancing the string tension.
While the interaction between MPIs and rope hadronisation mechanisms provides valuable insights on strange baryon production, discrepancies are observed in the data between baryons with different strangeness content, for instance \lmb and $\Xi$.

%% file: Chapters/conclusions.tex
\section{Conclusions}
\label{sec:conclusions}

This article presented a comprehensive multi-differential study of the production of \kzero strange mesons, $\Lambda$ strange baryons, and $\Xi$ multi-strange baryons in pp collisions at $\sqrt{s} = 13 ~\textrm{TeV}$ measured with the ALICE detector at the LHC. 
Using a novel approach, for the first time the production of strange hadrons at midrapidity is characterised not only as a function of the local particle multiplicity, but also as a function of the energy detected in ALICE Zero-Degree Calorimeters. 
Assuming the deposit in the calorimeters is proportional to the energy of collision remnants, the ZDCs provide an estimation of the effective energy available for particle production in the early stages of the collision. 

Transverse momentum spectra of \kzero, \lmb, and $\Xi$ are significantly affected by local multiplicity at fixed leading energy, whereas leading energy itself has a very limited effect on the \pt distribution once similar midrapidity multiplicities are selected.
Strangeness enhancement is observed with decreasing ZDC deposits, which implies a positive correlation of strange hadron production with the effective energy. This work complements previous studies on strange hadron production at midrapidity, by disentangling the correlation between the local charged-particle multiplicity and the ZDC energy through a multi-differential technique. Once the energy measured at forward rapidity is fixed, the \lmb and $\Xi$ enhancement with multiplicity is no longer observed. Instead, the strange baryon yield ratios show a mild anti-correlation with particle production at midrapidity. For similar midrapidity multiplicities, the production of \lmb and $\Xi$ strange baryons per charged particle is enhanced with increasing effective energy (anti-correlated with the ZDC energy). These effects are not observed for \kzero mesons, which display a very weak correlation with the effective energy. The largely different trends observed in this multi-differential analysis for $\Lambda$ and \kzero hadrons suggest that the production of strange baryons and mesons in pp collisions may be driven by different mechanisms. The enhancement of strange baryons appears to be strongly correlated with the effective energy; on the other hand, the hadronisation of strange quarks into \kzero mesons does not seem to be strongly influenced by the energy available in the early stages of the collision. It is worth noting that the dependence of the yield ratios to the $ZN$ energy differs between \lmb and $\Xi$ baryons, suggesting that the strangeness content plays a role in the observed behaviour. 

The higher integrated luminosity collected by ALICE in Run 3 will allow us to extend this novel approach to baryons with larger strangeness content, such as the $\Omega$, for which the statistical precision obtained with the Run 2 samples did not allow for a double-differential analysis.

The results are compared to Monte Carlo simulations based on two tunes of the PYTHIA 8 event generator: the Monash tune and the one including the improved colour reconnection and rope hadronisation mechanisms (QCD-CR+Ropes).
In general, the effective energy is observed to be strongly correlated with the number of MPI in PYTHIA. The data-model comparison of the predicted strange-particle production yields with the measured ones suggests that the interplay between the MPI and the CR+Ropes hadronisation mechanism is required to reproduce the observed enhancement of strange baryons $\Lambda$ and $\Xi$. However, when selecting events with similar effective energy these mechanisms fail to reproduce the measured multiplicity dependence of $\Lambda$ and $\Xi$ production.
No dynamics in terms of multiplicity and effective energy is predicted by both tunes for the production of \kzero mesons, suggesting that the CR+Ropes hadronisation mechanism results in an enhanced production of (strange) baryons, rather than of strangeness itself.

%% file: fa_2024-08-27_Opt_C.tex

The ALICE Collaboration would like to thank all its engineers and technicians for their invaluable contributions to the construction of the experiment and the CERN accelerator teams for the outstanding performance of the LHC complex.
The ALICE Collaboration gratefully acknowledges the resources and support provided by all Grid centres and the Worldwide LHC Computing Grid (WLCG) collaboration.
The ALICE Collaboration acknowledges the following funding agencies for their support in building and running the ALICE detector:
A. I. Alikhanyan National Science Laboratory (Yerevan Physics Institute) Foundation (ANSL), State Committee of Science and World Federation of Scientists (WFS), Armenia;
Austrian Academy of Sciences, Austrian Science Fund (FWF): [M 2467-N36] and Nationalstiftung f\"{u}r Forschung, Technologie und Entwicklung, Austria;
Ministry of Communications and High Technologies, National Nuclear Research Center, Azerbaijan;
Conselho Nacional de Desenvolvimento Cient\'{\i}fico e Tecnol\'{o}gico (CNPq), Financiadora de Estudos e Projetos (Finep), Funda\c{c}\~{a}o de Amparo \`{a} Pesquisa do Estado de S\~{a}o Paulo (FAPESP) and Universidade Federal do Rio Grande do Sul (UFRGS), Brazil;
Bulgarian Ministry of Education and Science, within the National Roadmap for Research Infrastructures 2020-2027 (object CERN), Bulgaria;
Ministry of Education of China (MOEC) , Ministry of Science \& Technology of China (MSTC) and National Natural Science Foundation of China (NSFC), China;
Ministry of Science and Education and Croatian Science Foundation, Croatia;
Centro de Aplicaciones Tecnol\'{o}gicas y Desarrollo Nuclear (CEADEN), Cubaenerg\'{\i}a, Cuba;
Ministry of Education, Youth and Sports of the Czech Republic, Czech Republic;
The Danish Council for Independent Research | Natural Sciences, the VILLUM FONDEN and Danish National Research Foundation (DNRF), Denmark;
Helsinki Institute of Physics (HIP), Finland;
Commissariat \`{a} l'Energie Atomique (CEA) and Institut National de Physique Nucl\'{e}aire et de Physique des Particules (IN2P3) and Centre National de la Recherche Scientifique (CNRS), France;
Bundesministerium f\"{u}r Bildung und Forschung (BMBF) and GSI Helmholtzzentrum f\"{u}r Schwerionenforschung GmbH, Germany;
General Secretariat for Research and Technology, Ministry of Education, Research and Religions, Greece;
National Research, Development and Innovation Office, Hungary;
Department of Atomic Energy Government of India (DAE), Department of Science and Technology, Government of India (DST), University Grants Commission, Government of India (UGC) and Council of Scientific and Industrial Research (CSIR), India;
National Research and Innovation Agency - BRIN, Indonesia;
Istituto Nazionale di Fisica Nucleare (INFN), Italy;
Japanese Ministry of Education, Culture, Sports, Science and Technology (MEXT) and Japan Society for the Promotion of Science (JSPS) KAKENHI, Japan;
Consejo Nacional de Ciencia (CONACYT) y Tecnolog\'{i}a, through Fondo de Cooperaci\'{o}n Internacional en Ciencia y Tecnolog\'{i}a (FONCICYT) and Direcci\'{o}n General de Asuntos del Personal Academico (DGAPA), Mexico;
Nederlandse Organisatie voor Wetenschappelijk Onderzoek (NWO), Netherlands;
The Research Council of Norway, Norway;
Pontificia Universidad Cat\'{o}lica del Per\'{u}, Peru;
Ministry of Science and Higher Education, National Science Centre and WUT ID-UB, Poland;
Korea Institute of Science and Technology Information and National Research Foundation of Korea (NRF), Republic of Korea;
Ministry of Education and Scientific Research, Institute of Atomic Physics, Ministry of Research and Innovation and Institute of Atomic Physics and Universitatea Nationala de Stiinta si Tehnologie Politehnica Bucuresti, Romania;
Ministry of Education, Science, Research and Sport of the Slovak Republic, Slovakia;
National Research Foundation of South Africa, South Africa;
Swedish Research Council (VR) and Knut \& Alice Wallenberg Foundation (KAW), Sweden;
European Organization for Nuclear Research, Switzerland;
Suranaree University of Technology (SUT), National Science and Technology Development Agency (NSTDA) and National Science, Research and Innovation Fund (NSRF via PMU-B B05F650021), Thailand;
Turkish Energy, Nuclear and Mineral Research Agency (TENMAK), Turkey;
National Academy of  Sciences of Ukraine, Ukraine;
Science and Technology Facilities Council (STFC), United Kingdom;
National Science Foundation of the United States of America (NSF) and United States Department of Energy, Office of Nuclear Physics (DOE NP), United States of America.
In addition, individual groups or members have received support from:
Czech Science Foundation (grant no. 23-07499S), Czech Republic;
FORTE project, reg.\ no.\ CZ.02.01.01/00/22\_008/0004632, Czech Republic, co-funded by the European Union, Czech Republic;
European Research Council (grant no. 950692), European Union;
ICSC - Centro Nazionale di Ricerca in High Performance Computing, Big Data and Quantum Computing, European Union - NextGenerationEU;
Academy of Finland (Center of Excellence in Quark Matter) (grant nos. 346327, 346328), Finland.

%% file: Alice_Authorlist_2024-08-27_Opt_C.tex
\begin{flushleft} 
\small

S.~Acharya\,\orcidlink{0000-0002-9213-5329}\,$^{\rm 127}$, 
A.~Agarwal$^{\rm 135}$, 
G.~Aglieri Rinella\,\orcidlink{0000-0002-9611-3696}\,$^{\rm 32}$, 
L.~Aglietta\,\orcidlink{0009-0003-0763-6802}\,$^{\rm 24}$, 
M.~Agnello\,\orcidlink{0000-0002-0760-5075}\,$^{\rm 29}$, 
N.~Agrawal\,\orcidlink{0000-0003-0348-9836}\,$^{\rm 25}$, 
Z.~Ahammed\,\orcidlink{0000-0001-5241-7412}\,$^{\rm 135}$, 
S.~Ahmad\,\orcidlink{0000-0003-0497-5705}\,$^{\rm 15}$, 
S.U.~Ahn\,\orcidlink{0000-0001-8847-489X}\,$^{\rm 71}$, 
I.~Ahuja\,\orcidlink{0000-0002-4417-1392}\,$^{\rm 37}$
% ...

\section*{Affiliation Notes}

$^{\rm I}$ Deceased\\
$^{\rm II}$ Also at: Max-Planck-Institut fur Physik, Munich, Germany\\
$^{\rm III}$ Also at: Italian National Agency for New Technologies, Energy and Sustainable Economic Development (ENEA), Bologna, Italy\\
$^{\rm IV}$ Also at: Dipartimento DET del Politecnico di Torino, Turin, Italy\\
$^{\rm V}$ Also at: Yildiz Technical University, Istanbul, T\"{u}rkiye\\
$^{\rm VI}$ Also at: Department of Applied Physics, Aligarh Muslim University, Aligarh, India\\
$^{\rm VII}$ Also at: Institute of Theoretical Physics, University of Wroclaw, Poland\\
$^{\rm VIII}$ Also at: An institution covered by a cooperation agreement with CERN\\

\section*{Collaboration Institutes}

$^{1}$ A.I. Alikhanyan National Science Laboratory (Yerevan Physics Institute) Foundation, Yerevan, Armenia\\
$^{2}$ AGH University of Krakow, Cracow, Poland\\
$^{3}$ Bogolyubov Institute for Theoretical Physics, National Academy of Sciences of Ukraine, Kiev, Ukraine\\
$^{4}$ Bose Institute, Department of Physics  and Centre for Astroparticle Physics and Space Science (CAPSS), Kolkata, India\\
$^{5}$ California Polytechnic State University, San Luis Obispo, California, United States\\
$^{6}$ Central China Normal University, Wuhan, China\\
$^{7}$ Centro de Aplicaciones Tecnol\'{o}gicas y Desarrollo Nuclear (CEADEN), Havana, Cuba\\
$^{8}$ Centro de Investigaci\'{o}n y de Estudios Avanzados (CINVESTAV), Mexico City and M\'{e}rida, Mexico\\
$^{9}$ Chicago State University, Chicago, Illinois, United States\\
$^{10}$ China Institute of Atomic Energy, Beijing, China\\
$^{11}$ China University of Geosciences, Wuhan, China\\
$^{12}$ Chungbuk National University, Cheongju, Republic of Korea\\
$^{13}$ Comenius University Bratislava, Faculty of Mathematics, Physics and Informatics, Bratislava, Slovak Republic\\
$^{14}$ Creighton University, Omaha, Nebraska, United States\\
$^{15}$ Department of Physics, Aligarh Muslim University, Aligarh, India\\
$^{16}$ Department of Physics, Pusan National University, Pusan, Republic of Korea\\
$^{17}$ Department of Physics, Sejong University, Seoul, Republic of Korea\\
$^{18}$ Department of Physics, University of California, Berkeley, California, United States\\
$^{19}$ Department of Physics, University of Oslo, Oslo, Norway\\
$^{20}$ Department of Physics and Technology, University of Bergen, Bergen, Norway\\
$^{21}$ Dipartimento di Fisica, Universit\`{a} di Pavia, Pavia, Italy\\
$^{22}$ Dipartimento di Fisica dell'Universit\`{a} and Sezione INFN, Cagliari, Italy\\
$^{23}$ Dipartimento di Fisica dell'Universit\`{a} and Sezione INFN, Trieste, Italy\\
$^{24}$ Dipartimento di Fisica dell'Universit\`{a} and Sezione INFN, Turin, Italy\\
$^{25}$ Dipartimento di Fisica e Astronomia dell'Universit\`{a} and Sezione INFN, Bologna, Italy\\
$^{26}$ Dipartimento di Fisica e Astronomia dell'Universit\`{a} and Sezione INFN, Catania, Italy\\
$^{27}$ Dipartimento di Fisica e Astronomia dell'Universit\`{a} and Sezione INFN, Padova, Italy\\
$^{28}$ Dipartimento di Fisica `E.R.~Caianiello' dell'Universit\`{a} and Gruppo Collegato INFN, Salerno, Italy\\
$^{29}$ Dipartimento DISAT del Politecnico and Sezione INFN, Turin, Italy\\
$^{30}$ Dipartimento di Scienze MIFT, Universit\`{a} di Messina, Messina, Italy\\
$^{31}$ Dipartimento Interateneo di Fisica `M.~Merlin' and Sezione INFN, Bari, Italy\\
$^{32}$ European Organization for Nuclear Research (CERN), Geneva, Switzerland\\
$^{33}$ Faculty of Electrical Engineering, Mechanical Engineering and Naval Architecture, University of Split, Split, Croatia\\
$^{34}$ Faculty of Engineering and Science, Western Norway University of Applied Sciences, Bergen, Norway\\
$^{35}$ Faculty of Nuclear Sciences and Physical Engineering, Czech Technical University in Prague, Prague, Czech Republic\\
$^{36}$ Faculty of Physics, Sofia University, Sofia, Bulgaria\\
$^{37}$ Faculty of Science, P.J.~\v{S}af\'{a}rik University, Ko\v{s}ice, Slovak Republic\\
$^{38}$ Frankfurt Institute for Advanced Studies, Johann Wolfgang Goethe-Universit\"{a}t Frankfurt, Frankfurt, Germany\\
$^{39}$ Fudan University, Shanghai, China\\
$^{40}$ Gangneung-Wonju National University, Gangneung, Republic of Korea\\
$^{41}$ Gauhati University, Department of Physics, Guwahati, India\\
$^{42}$ Helmholtz-Institut f\"{u}r Strahlen- und Kernphysik, Rheinische Friedrich-Wilhelms-Universit\"{a}t Bonn, Bonn, Germany\\
$^{43}$ Helsinki Institute of Physics (HIP), Helsinki, Finland\\
$^{44}$ High Energy Physics Group,  Universidad Aut\'{o}noma de Puebla, Puebla, Mexico\\
$^{45}$ Horia Hulubei National Institute of Physics and Nuclear Engineering, Bucharest, Romania\\
$^{46}$ HUN-REN Wigner Research Centre for Physics, Budapest, Hungary\\
$^{47}$ Indian Institute of Technology Bombay (IIT), Mumbai, India\\
$^{48}$ Indian Institute of Technology Indore, Indore, India\\
$^{49}$ INFN, Laboratori Nazionali di Frascati, Frascati, Italy\\
$^{50}$ INFN, Sezione di Bari, Bari, Italy\\
$^{51}$ INFN, Sezione di Bologna, Bologna, Italy\\
$^{52}$ INFN, Sezione di Cagliari, Cagliari, Italy\\
$^{53}$ INFN, Sezione di Catania, Catania, Italy\\
$^{54}$ INFN, Sezione di Padova, Padova, Italy\\
$^{55}$ INFN, Sezione di Pavia, Pavia, Italy\\
$^{56}$ INFN, Sezione di Torino, Turin, Italy\\
$^{57}$ INFN, Sezione di Trieste, Trieste, Italy\\
$^{58}$ Inha University, Incheon, Republic of Korea\\
$^{59}$ Institute for Gravitational and Subatomic Physics (GRASP), Utrecht University/Nikhef, Utrecht, Netherlands\\
$^{60}$ Institute of Experimental Physics, Slovak Academy of Sciences, Ko\v{s}ice, Slovak Republic\\
$^{61}$ Institute of Physics, Homi Bhabha National Institute, Bhubaneswar, India\\
$^{62}$ Institute of Physics of the Czech Academy of Sciences, Prague, Czech Republic\\
$^{63}$ Institute of Space Science (ISS), Bucharest, Romania\\
$^{64}$ Institut f\"{u}r Kernphysik, Johann Wolfgang Goethe-Universit\"{a}t Frankfurt, Frankfurt, Germany\\
$^{65}$ Instituto de Ciencias Nucleares, Universidad Nacional Aut\'{o}noma de M\'{e}xico, Mexico City, Mexico\\
$^{66}$ Instituto de F\'{i}sica, Universidade Federal do Rio Grande do Sul (UFRGS), Porto Alegre, Brazil\\
$^{67}$ Instituto de F\'{\i}sica, Universidad Nacional Aut\'{o}noma de M\'{e}xico, Mexico City, Mexico\\
$^{68}$ iThemba LABS, National Research Foundation, Somerset West, South Africa\\
$^{69}$ Jeonbuk National University, Jeonju, Republic of Korea\\
$^{70}$ Johann-Wolfgang-Goethe Universit\"{a}t Frankfurt Institut f\"{u}r Informatik, Fachbereich Informatik und Mathematik, Frankfurt, Germany\\
$^{71}$ Korea Institute of Science and Technology Information, Daejeon, Republic of Korea\\
$^{72}$ KTO Karatay University, Konya, Turkey\\
$^{73}$ Laboratoire de Physique Subatomique et de Cosmologie, Universit\'{e} Grenoble-Alpes, CNRS-IN2P3, Grenoble, France\\
$^{74}$ Lawrence Berkeley National Laboratory, Berkeley, California, United States\\
$^{75}$ Lund University Department of Physics, Division of Particle Physics, Lund, Sweden\\
$^{76}$ Nagasaki Institute of Applied Science, Nagasaki, Japan\\
$^{77}$ Nara Women{'}s University (NWU), Nara, Japan\\
$^{78}$ National and Kapodistrian University of Athens, School of Science, Department of Physics , Athens, Greece\\
$^{79}$ National Centre for Nuclear Research, Warsaw, Poland\\
$^{80}$ National Institute of Science Education and Research, Homi Bhabha National Institute, Jatni, India\\
$^{81}$ National Nuclear Research Center, Baku, Azerbaijan\\
$^{82}$ National Research and Innovation Agency - BRIN, Jakarta, Indonesia\\
$^{83}$ Niels Bohr Institute, University of Copenhagen, Copenhagen, Denmark\\
$^{84}$ Nikhef, National institute for subatomic physics, Amsterdam, Netherlands\\
$^{85}$ Nuclear Physics Group, STFC Daresbury Laboratory, Daresbury, United Kingdom\\
$^{86}$ Nuclear Physics Institute of the Czech Academy of Sciences, Husinec-\v{R}e\v{z}, Czech Republic\\
$^{87}$ Oak Ridge National Laboratory, Oak Ridge, Tennessee, United States\\
$^{88}$ Ohio State University, Columbus, Ohio, United States\\
$^{89}$ Physics department, Faculty of science, University of Zagreb, Zagreb, Croatia\\
$^{90}$ Physics Department, Panjab University, Chandigarh, India\\
$^{91}$ Physics Department, University of Jammu, Jammu, India\\
$^{92}$ Physics Program and International Institute for Sustainability with Knotted Chiral Meta Matter (SKCM2), Hiroshima University, Hiroshima, Japan\\
$^{93}$ Physikalisches Institut, Eberhard-Karls-Universit\"{a}t T\"{u}bingen, T\"{u}bingen, Germany\\
$^{94}$ Physikalisches Institut, Ruprecht-Karls-Universit\"{a}t Heidelberg, Heidelberg, Germany\\
$^{95}$ Physik Department, Technische Universit\"{a}t M\"{u}nchen, Munich, Germany\\
$^{96}$ Politecnico di Bari and Sezione INFN, Bari, Italy\\
$^{97}$ Research Division and ExtreMe Matter Institute EMMI, GSI Helmholtzzentrum f\"ur Schwerionenforschung GmbH, Darmstadt, Germany\\
$^{98}$ Saga University, Saga, Japan\\
$^{99}$ Saha Institute of Nuclear Physics, Homi Bhabha National Institute, Kolkata, India\\
$^{100}$ School of Physics and Astronomy, University of Birmingham, Birmingham, United Kingdom\\
$^{101}$ Secci\'{o}n F\'{\i}sica, Departamento de Ciencias, Pontificia Universidad Cat\'{o}lica del Per\'{u}, Lima, Peru\\
$^{102}$ Stefan Meyer Institut f\"{u}r Subatomare Physik (SMI), Vienna, Austria\\
$^{103}$ SUBATECH, IMT Atlantique, Nantes Universit\'{e}, CNRS-IN2P3, Nantes, France\\
$^{104}$ Sungkyunkwan University, Suwon City, Republic of Korea\\
$^{105}$ Suranaree University of Technology, Nakhon Ratchasima, Thailand\\
$^{106}$ Technical University of Ko\v{s}ice, Ko\v{s}ice, Slovak Republic\\
$^{107}$ The Henryk Niewodniczanski Institute of Nuclear Physics, Polish Academy of Sciences, Cracow, Poland\\
$^{108}$ The University of Texas at Austin, Austin, Texas, United States\\
$^{109}$ Universidad Aut\'{o}noma de Sinaloa, Culiac\'{a}n, Mexico\\
$^{110}$ Universidade de S\~{a}o Paulo (USP), S\~{a}o Paulo, Brazil\\
$^{111}$ Universidade Estadual de Campinas (UNICAMP), Campinas, Brazil\\
$^{112}$ Universidade Federal do ABC, Santo Andre, Brazil\\
$^{113}$ Universitatea Nationala de Stiinta si Tehnologie Politehnica Bucuresti, Bucharest, Romania\\
$^{114}$ University of Cape Town, Cape Town, South Africa\\
$^{115}$ University of Derby, Derby, United Kingdom\\
$^{116}$ University of Houston, Houston, Texas, United States\\
$^{117}$ University of Jyv\"{a}skyl\"{a}, Jyv\"{a}skyl\"{a}, Finland\\
$^{118}$ University of Kansas, Lawrence, Kansas, United States\\
$^{119}$ University of Liverpool, Liverpool, United Kingdom\\
$^{120}$ University of Science and Technology of China, Hefei, China\\
$^{121}$ University of South-Eastern Norway, Kongsberg, Norway\\
$^{122}$ University of Tennessee, Knoxville, Tennessee, United States\\
$^{123}$ University of the Witwatersrand, Johannesburg, South Africa\\
$^{124}$ University of Tokyo, Tokyo, Japan\\
$^{125}$ University of Tsukuba, Tsukuba, Japan\\
$^{126}$ Universit\"{a}t M\"{u}nster, Institut f\"{u}r Kernphysik, M\"{u}nster, Germany\\
$^{127}$ Universit\'{e} Clermont Auvergne, CNRS/IN2P3, LPC, Clermont-Ferrand, France\\
$^{128}$ Universit\'{e} de Lyon, CNRS/IN2P3, Institut de Physique des 2 Infinis de Lyon, Lyon, France\\
$^{129}$ Universit\'{e} de Strasbourg, CNRS, IPHC UMR 7178, F-67000 Strasbourg, France, Strasbourg, France\\
$^{130}$ Universit\'{e} Paris-Saclay, Centre d'Etudes de Saclay (CEA), IRFU, D\'{e}partment de Physique Nucl\'{e}aire (DPhN), Saclay, France\\
$^{131}$ Universit\'{e}  Paris-Saclay, CNRS/IN2P3, IJCLab, Orsay, France\\
$^{132}$ Universit\`{a} degli Studi di Foggia, Foggia, Italy\\
$^{133}$ Universit\`{a} del Piemonte Orientale, Vercelli, Italy\\
$^{134}$ Universit\`{a} di Brescia, Brescia, Italy\\
$^{135}$ Variable Energy Cyclotron Centre, Homi Bhabha National Institute, Kolkata, India\\
$^{136}$ Warsaw University of Technology, Warsaw, Poland\\
$^{137}$ Wayne State University, Detroit, Michigan, United States\\
$^{138}$ Yale University, New Haven, Connecticut, United States\\
$^{139}$ Yonsei University, Seoul, Republic of Korea\\
$^{140}$ Affiliated with an institute covered by a cooperation agreement with CERN\\
$^{141}$ Affiliated with an international laboratory covered by a cooperation agreement with CERN.\\

\end{flushleft} 